\documentclass[11pt]{article}
\usepackage[utf8]{inputenc}
\usepackage{caption}
\usepackage{subcaption}
\usepackage{braket,slashed,bm}
\usepackage{jheppub}
\usepackage{simplewick}
\usepackage{array,multirow}
\usepackage{amsmath} \DeclareMathAlphabet{\mathpzc}{OT1}{pzc}{m}{it}
\usepackage[normalem]{ulem}
\usepackage{calligra}
\usepackage{floatrow}
\DeclareMathAlphabet{\mathpzc}{OT1}{pzc}{m}{it}
\usepackage{mathrsfs}
\usepackage{tikz}
\usepackage{rotating}

\usepackage{pgf}

\usepackage{subcaption}
\usepackage{lscape}
\usepackage{hyperref}
\usepackage{graphicx}
\usepackage{booktabs}
\newfloatcommand{capbtabbox}{table}[][\FBwidth]

\graphicspath{{./Figures/}}

\newcommand{\CLASS}{\texttt{CLASS}}

\newcommand{\Healpix}{\texttt{HEALPix}}

\newcommand{\Quicklens}{\texttt{QuickLens }}

\def\beq{\begin{equation}}
\def\eeq{\end{equation}}
\def\bea{\begin{eqnarray}}
\def\eea{\end{eqnarray}}

\newcommand{\mpl}{M_{\rm pl}}

\pgfkeys{/pgf/number format/.cd ,precision=2,sci generic={exponent={\times 10^{#1}}}}

\allowdisplaybreaks

\begin{document}

\title{
Probing Cosmological Particle Production and Pairwise Hotspots with Deep Neural Networks
}
\author[a]{Taegyun Kim,}
\author[b,c,d]{Jeong Han Kim,}
\author[e,f]{Soubhik Kumar,}
\author[a]{Adam Martin,}
\author[g]{Moritz M{\"u}nchmeyer,}
\author[a]{Yuhsin Tsai}

\affiliation[a]{Department of Physics, University of Notre Dame, IN 46556, USA}
\affiliation[b]{Department of Physics, Chungbuk National University, Cheongju, Chungbuk 28644, Korea}
\affiliation[c]{Center for Theoretical Physics of the Universe,
Institute for Basic Science, Daejeon 34126, Korea}
\affiliation[d]{Korea Institute for Advanced Study (KIAS), School of Physics, Seoul 02455, Korea}
\affiliation[e]{Berkeley Center for Theoretical Physics, Department of Physics,
University of California, Berkeley, CA 94720, USA}
\affiliation[f]{Theoretical Physics Group, Lawrence Berkeley National Laboratory, Berkeley, CA 94720, USA}
\affiliation[g]{Department of Physics, University of Wisconsin-Madison, 1150 University Ave, Madison, WI 53706, USA}
\emailAdd{tkim12@nd.edu}
\emailAdd{jeonghan.kim@cbu.ac.kr}
\emailAdd{soubhik@berkeley.edu}
\emailAdd{amarti41@nd.edu}
\emailAdd{muenchmeyer@wisc.edu}
\emailAdd{ytsai3@nd.edu}

\abstract{
Particles with masses much larger than the inflationary Hubble scale, $H_I$, can be pair-produced non-adiabatically during inflation.
Due to their large masses, the produced particles modify the curvature perturbation around their locations.
These localized perturbations eventually give rise to  localized signatures on the Cosmic Microwave Background (CMB), in particular, pairwise hotspots (PHS). 
In this work, we show that Convolutional Neural Networks (CNN) provide a powerful tool for identifying PHS on the CMB.
While for a given hotspot profile a traditional Matched Filter Analysis is known to be optimal, a Neural Network learns to effectively detect the large variety of shapes that can arise in realistic models of particle production. 
Considering an idealized situation where the dominant background to the PHS signal comes from the standard CMB fluctuations, we show that a CNN can isolate the PHS with $\mathcal{O}(10)\%$ efficiency even if the hotspot temperature is $\mathcal{O}(10)$ times smaller than the average CMB fluctuations.
Overall, the CNN search is sensitive to heavy particle masses $M_0/H_I=\mathcal{O}(200)$, and constitutes one of the unique probes of very high energy particle physics.
}
\maketitle

\section{Introduction}
An era of cosmic inflation~\cite{Guth:1980zm, Linde:1981mu, Albrecht:1982wi} in the primordial Universe remains an attractive paradigm to explain the origin of (approximately) scale invariant, Gaussian, and adiabatic primordial perturbations, inferred through cosmic microwave background (CMB) and large scale structure (LSS) observations.
This inflationary era can be characterized by a rapid expansion of spacetime, controlled by an approximately constant Hubble scale $H_I$.
Excitingly, based on the current constraints, $H_I$ can be as large as $5\times 10^{13}$~GeV~\cite{Planck:2018jri}.
This fact, coupled with the feature that particles with masses up to order $H_I$ can get quantum mechanically produced during inflation, makes the inflationary era a natural and unique arena to {\it directly} probe very high energy particle physics.

There are several classes of mechanisms through which heavy particles, which we label as $\chi$, can be produced during inflation. 
When their mass $m_\chi \lesssim H_I$, quantum fluctuations of the inflationary spacetime itself can efficiently produce the $\chi$ particles.
However, for $m_\chi \gg H_I$, this production gets suppressed exponentially as $e^{-\pi m_\chi /H_I}$~\cite{Birrell:1982ix}, and other mechanisms are necessary for efficient particle production to occur.

To illustrate this,  we consider the standard slow-roll inflationary paradigm containing an inflaton field $\phi$ whose homogeneous component we denote by  $\phi_0(t)$.
Normalization of the primordial scalar power spectrum requires the `kinetic energy' of this homogeneous component to be $|d{\phi}_0/dt|^{1/2} \approx 60 H_I$~\cite{Planck:2018jri}.
Therefore, heavy particles, if appropriately coupled to the inflaton kinetic term, can be efficiently produced for $m_\chi\lesssim 60 H_I $.
One class of examples of this involve a coupling of the type $\partial_\mu\phi J^\mu$ where $J^\mu$ is a charged current made up of the $\chi$ field.
For some recent work implementing this idea see, e.g, Refs.~\cite{  Adshead:2015kza, Adshead:2018oaa,
Chen:2018xck, Wang:2019gbi, Hook:2019zxa, Wang:2020ioa, Bodas:2020yho, Sou:2021juh, Chen:2023txq}. 
In these constructions, heavy particle production happens continuously in time, in a scale-invariant fashion.
In other words, the coupling of the inflaton to $\chi$ particles does not break the shift symmetry, $\phi\rightarrow \phi+{\rm constant}$, of the inflaton.

A different class of mechanisms can lead to particle production at specific times during the inflationary evolution.
This can happen if the shift symmetry of the inflaton is broken in a controlled manner, e.g. to a discrete shift symmetry.
This breaking of shift symmetry translates into a violation of scale invariance, and selects out specific time instant(s) when particle production can occur.
Examples of such mechanisms appear in Refs.~\cite{ Chung:1998zb, Chung:1999ve, Kofman:2004yc,Romano:2008rr, Barnaby:2009mc, Flauger:2016idt}, and see Refs.~\cite{Chen:2010xka, Chluba:2015bqa} for reviews.

A particularly interesting example of this latter mechanism arises in the context of ultra-heavy particles with time-dependent masses.
More specifically, suppose $m_\chi$ varies as a function of $\phi$ in a way such that, as $\phi$ passes through a specific point $\phi_*$  on the inflaton potential at time $t_*$, $m_\chi(\phi)$ passes through a local minimum. 
In this case, non-adiabatic $\chi$ particle production can occur at time $t_*$.
Following their production, $\chi$ particles can again become heavy, $m_\chi \gg |d{\phi}_0/dt|^{1/2}$, and owing to this large mass they can backreact on the inflationary spacetime, contributing to the curvature perturbation around their locations.

We can describe the effects of these additional curvature perturbations qualitatively in the following way, leaving the details for the next section.
Following their production, the perturbations exit the horizon when their wavelengths become larger than $1/H_I$ and become frozen in time.
After the end of inflation, they eventually reenter the horizon and source additional under- or over-densities in the thermal plasma in the radiation dominated Universe.
Overdense regions, for example, would trap more plasma, and therefore would emit more photons at the time of CMB decoupling.\footnote{To be more accurate, one also need to take into account the gravitational redshift of the photons as they climb out of the gravitational potential wells. We will compute this effect in the next section.} 
Therefore, we would observe localized regions on the sky where CMB would appear hotter than usual.
As we will discuss below, the sizes of these localized `spots' are determined by the size of the comoving horizon, $\eta_*$, at the time of particle production $t_*$.
While $\eta_*$ can take any value, for concreteness we will consider $\eta_* \sim 100$~Mpc in this work. 
This implies that the localized spots would subtend $\sim 1^\circ$ on the CMB sky.

The next question one may ask is what is an efficient strategy to look for such signatures.
Since this scenario is associated with a violation of scale invariance, characterized by $\eta_*$, one would expect to see `features' on the CMB power spectrum or even higher-point correlation functions.
However, in the regime we focus on, the total number of produced $\chi$ particles is still small to the extent that the CMB power spectrum is minimally affected, as we explicitly check later.
On the other hand, the spots can still be individually bright enough such that we can look for them directly in position space.
Indeed, this class of signatures in the context of heavy particle production were discussed in Refs.~\cite{Fialkov:2009xm, Maldacena:2015bha}, and in Ref.~\cite{Kim:2021ida} the associated CMB phenomenology was described and a simple `cut-and-count' search strategy was developed. 
Using the cut-and-count strategy, Ref.~\cite{Kim:2021ida} constrained the parameter space of ultra-heavy scalars and illustrated regions where a position space search is more powerful than power spectrum-based searches.

In more detail, Ref.~\cite{Kim:2021ida} considered a single instance of particle production during the time when CMB-observable modes exit the horizon.
Conservation of momentum implies that such heavy particles are produced in pairs.
However, owing to their large mass, the particles do not drift significantly following their production, and it was argued that the separation between the two particles forming a pair can be taken to be a uniformly random number between 0 and $\eta_*$.
Finally, it was shown that the coupling $g$ of $\chi$ to the inflaton determines how hot/cold the associated spot on the CMB is with the heavy particle mass $m_\chi$ determining the total number of such spots on the sky.
To summarize, the three parameters determining the hot/cold spot phenomenology are $\{g, m_\chi, \eta_*\}$, as will be reviewed in more detail in the next section.
While both cold or hot spots can arise depending on the value of $\eta_*$, for the choices of $\eta_*$ in this work, only hotspots will appear on the CMB. Therefore, we will often be referring to these localized spots as hotspots, in particular as pairwise hotspots (PHS) since the spots appear in pairs.
 
In the present work, we improve upon Ref.~\cite{Kim:2021ida} in several important ways. First, in Ref.~\cite{Kim:2021ida} we only considered hotspots that lie within the last scattering surface, with a thickness of $\Delta\eta \approx 19$~Mpc~\cite{Hadzhiyska:2018mwh}.
In this work we adopt a more realistic setup and include hotspots that are distributed in a larger region around the last scattering surface.
We take this region to have a thickness of $2\eta_*$ and we show in Sec.~\ref{sec.PHS} how hotspots lying outside the $\Delta\eta$ shell can still affect the CMB.
The overall signature of PHS then changes non-trivially.
For instance, with the improved treatment we can have one spot of a pair lying on the CMB surface, while the other can lie off the CMB surface, leading to an asymmetric signal.

Second, we develop a neural network (NN)-based search for the hotspot profiles. In principle, a neural network is not necessary to search for a profile of known shape which is linearly added to the Gaussian background. In this case, the standard method of constructing a so-called matched filter can be shown to be the optimal statistic to detect the profile (see, e.g.,~\cite{Munchmeyer:2019wlh}). Matched filter-based searches for radially symmetric profiles in the CMB have been previously reported for example in~\cite{Osborne:2013jea,Osborne:2013hea,McEwen:2012uk}, with the physical motivation of searching for inflationary bubble collisions. Various matched filters have also been used in the Planck Anisotropy and Statistics Analysis \cite{Planck:2015igc,Planck:2019evm} without finding a significant excess. However, the signal which we are looking for here is more complicated. Profiles come in pairs (breaking radial symmetry of the profile), they can be overlapping, and, depending on their production time and orientation with respect to the surface of last scattering, their appearance on the CMB changes. While it is in principle possible to cover the entire space of profiles with a very large bank of matched filters, this would be a complicated and computationally challenging approach. A neural network, on the other hand, can learn an effective representation of these filters which interpolates well between all profile shapes, including overlapping ones. We also implement the matched filter method below, and show that in the simplified case with a single profile type, our neural network performs similar to the optimal matched filter. 
 
This work is organized as follows. 
We first describe a simple model of $\chi$ particle production in Sec.~\ref{sec.PHS} and summarize how the total number of produced particles depends on the model parameters along with various properties of the PHS.
We improve the calculation of the hotspot profiles by taking into account the line-of-sight distance to the location of the hotspots which can be off the CMB surface.
In Sec.~\ref{sec:image}, we describe the simulation of the PHS signals and the CMB maps in angular space, assuming that the dominant background to the PHS signal comes from the standard CMB fluctuations. In Sec.~\ref{sec.CNN}, we describe the convolutional neural network (CNN) analysis and estimate the sensitivity the CNN can achieve for a PHS search. We then translate this sensitivity to the mass-coupling parameter space of the heavy particles. We also compare the CNN analysis with a matched filter analysis for simplified hotspot configurations. We conclude in Sec.~\ref{sec.concl}.

\section{Pairwise Hotspot Signals}\label{sec.PHS}
To model heavy particle production, we consider a scenario where the mass of $\chi$ is inflaton-dependent, $m_\chi(\phi)$.
Therefore as $\phi$ moves along its potential, efficient, non-adiabatic particle production can occur if $m_\chi(\phi)$ varies with $\phi$ rapidly.
With a mass term $m_\chi(\phi)^2\chi^2$, {\it pairs} of $\chi$ particles would be produced, as required by three-momenta conservation.
The phenomenology of such heavy particles depend on their mass, coupling to the inflaton, and the horizon size at the time of their production. We now review these properties more qualitatively, referring to Ref.~\cite{Kim:2021ida} for a more complete discussion.

\subsection{Inflationary Particle Production}
We parametrize the inflationary spacetime metric as,
\begin{align}
    ds^2 = -dt^2 + a^2(t) d\vec{x}^2,
\end{align}
with the scale factor $a(t) = e^{H_I t}$ and $H_I$ the Hubble scale during inflation that we take to be (approximately) constant. 
To model particle production in a simple way, we assume $m_\chi(\phi)$ passes through a minimum as $\phi$ crosses a field value $\phi_*$.
Then we can expand $m_\chi(\phi)$ near $\phi_*$ as,
\begin{align}
    m_\chi(\phi) = m_\chi(\phi_*) + \frac{1}{2}m''_\chi(\phi_*)(\phi-\phi_*)^2 +\cdots,
\end{align}
where primes denote derivatives with respect to $\phi$.
Thus the mass term would appear in the potential as,
\begin{align}
    m_\chi(\phi)^2\chi^2 = m_\chi(\phi_*)^2\chi^2 + m_\chi(\phi_*)m''_\chi(\phi_*)(\phi-\phi_*)^2\chi^2 + \cdots.
\end{align}
While away from $\phi_*$, $m_\chi(\phi)$ can vary in different ways, most of the important features of particle production are determined by the behavior of $m_\chi(\phi)$ around $\phi_*$.
For example, the number density of $\chi$ particles is determined by $m_\chi(\phi_*)$, as we will see below.
Similarly, the spatial profiles of the hotspots on the CMB is determined by the dependence $(\phi-\phi_*)^2 \sim \dot{\phi}_0^2(t-t_*)^2 \sim (\dot{\phi}_0/H_I)^2\log(\eta/\eta_*)^2$, where we have used the relation between $t$ and conformal time $\eta$, $\eta = (-1/H_I)e^{-H_It}$ (an overdot here denotes a derivative with respect to time).
Given the importance of the physics around $\phi_*$, we will denote, $m_\chi(\phi_*)^2 \equiv M_0^2$, $m_\chi(\phi_*)m''_\chi(\phi_*) \equiv g^2$, and $\phi_* \equiv \mu/g$, to 
describe particle production.
Thus we will write the Lagrangian for $\chi$ as,
\begin{align}
\label{eq:lag}
    \mathcal{L}_\chi = -\frac{1}{2}(\partial_\mu \chi)^2 - \frac{1}{2} \left((g \phi - \mu)^2 + M_0^2\right) \chi^2.
\end{align}
As $\phi$ nears the field value $\phi_*$, the mass of the $\chi$ field changes non-adiabatically and particle production can occur.

The efficiency of particle production depends on the parameters $g$, $M_0$, and $\eta_*$, the size of the comoving horizon at the time of particle production.
This can be computed using the standard Bogoliubov approach, and resulting probability of particle production is given by~\cite{Kofman:1997yn,Flauger:2016idt},
\begin{align}\label{eq:bog_beta}
	|\beta|^2 = \exp\left(-\frac{\pi(M_0^2-2H_I^2+k^2\eta_*^2 H_I^2)}{g|\dot{\phi}_0|}\right). 
\end{align}
The normalization of the scalar primordial power spectrum, in the context of single-field slow-roll inflation, fixes $A_s = H_I^4/(4\pi^2\dot{\phi}_0^2) \approx 2.1 \times 10^{-9}$~\cite{Planck:2018jri} which determines $\dot{\phi}_0 \approx (58.9 H_I)^2$.

The above expression~\eqref{eq:bog_beta} characterizes the probability of particle production with physical momentum $k_p=k\eta_* H_I$.
The total number density of particles can then be computed by integrating over all such $k$-modes,
\begin{align}\label{eq.nintgl}
	n = \frac{1}{2\pi^2} \int_0^\infty dk_{p} k_{p}^2 e^{-\pi k_{p}^2/(g|\dot{\phi}_0|)} e^{-\pi (M_0^2-2H_I^2)/(g|\dot{\phi}_0|)} = \frac{1}{8\pi^3}\left(g \dot{\phi}_0\right)^{3/2} e^{-\pi (M_0^2-2H_I^2)/(g|\dot{\phi}_0|)}.
\end{align}
From an observational perspective, it is more convenient to relate $n$ to the total number of spots that would be visible on the CMB sky.
To that end, we need to specify the associated spacetime volume.
Considering a shell of thickness $\Delta \eta_s$ around the CMB surface, the total number of spots in that shell is given by~\cite{Kim:2021ida},
\begin{equation}
\begin{aligned}\label{eq.Nspots}
	N_{\rm spots} & = n \times \left(\frac{a_*}{a_0}\right)^3\times 4\pi \chi_{\rm rec}^2\Delta \eta_s\,,\\
 & = \frac{1}{2\pi^2} \left(\frac{g \dot{\phi}_0}{H_I^2}\right)^{3/2} \frac{\Delta \eta_s}{\chi_{\rm rec}}(k_*\chi_{\rm rec})^3 e^{-\pi (M_0^2-2H_I^2)/(g|\dot{\phi}_0|)}\,,\\
     & \approx 4 \times 10^8 \times g^{3/2} \left(\frac{\Delta \eta_s}{100~{\rm Mpc}}\right) \left(\frac{100~{\rm Mpc}}{\eta_*}\right)^3 e^{-\pi (M_0^2-2H_I^2)/(g|\dot{\phi}_0|)}\,.
\end{aligned}
\end{equation}
Here $a_*$ and $a_0 = 1$ are the scale factors at the time of particle production and today, respectively. 
The quantity $\chi_{\rm rec}$ is the distance of the CMB surface from us and approximately equals $13871$~Mpc, obtained from Planck's best-fit $\Lambda$CDM parameters, and $k_* = a_* H_I = 1/\eta_*$ is the mode that exits the horizon at the time of particle production.

\subsection{Effect on the CMB}
We now discuss the detailed properties of the spots and how they modify the CMB.
\paragraph{Primordial Curvature Perturbation from Heavy Particles.}
Owing to their large mass, the heavy particles can backreact on the spacetime metric around their locations, and can give rise to non-trivial curvature perturbations.
The profile of such a curvature perturbation can be computed using the in-in formalism and the result is given by~\cite{Maldacena:2015bha}, 
\begin{equation}\label{eq:hsprofile}
\langle \zeta_{\rm HS}(r)\rangle=\frac{H_I}{8\epsilon \pi M_{\rm pl}^2}\begin{cases} M(\eta=-r), & \mbox{if } r\leq\eta_*\\ 0, & \mbox{if } r>\eta_* \end{cases}\,.
\end{equation}
Here $\epsilon = |\dot{H}_I|/H_I^2$ is a slow-roll parameter, and we have anticipated that this curvature perturbation would give rise to a hotspot (HS), rather than a coldspot.
Importantly, the variation of the mass as a function of conformal time $\eta$ controls the spatial profile.
This variation can be computed from Eq.~\eqref{eq:lag} by noting the slow-roll equation $\phi - \phi_* \approx \dot{\phi}_0(t-t_*)$, which gives
\begin{align}
\label{eq:time_dep_mass}
    M(\eta)^2 = \frac{g^2\dot{\phi}_0^2}{H_I^2}\ln(\eta/\eta_*)^2 + M_0^2.
\end{align}
 Here we have used the relation between cosmic time $t$ and the conformal time $\eta$, that also determines the size of the comoving horizon, $t - t_* = -(1/H_I)\ln\left(\eta/\eta_*\right)$. 

Using the slow-roll relation $\dot{\phi}_0^2 = 2 \epsilon H_I^2 \mpl^2$ and the fact that $M_0^2 \sim g |\dot{\phi}_0|$ so that $N_{\rm spots}$ is not significantly exponentially suppressed (see Eq.~\eqref{eq.Nspots}), we can drop the contribution of the second term in Eq.~\eqref{eq:time_dep_mass} away from $\eta_*$. 
The profile can then be simply written as,
\begin{align}\label{eq:hs_simple}
    \langle\zeta_{\rm HS}(r)\rangle = \frac{g H^2}{4\pi |\dot{\phi}_0|}\ln(\eta_*/r) \theta(\eta_* - r).
\end{align}
Given the typical size of a standard quantum mechanical fluctuation $\langle \zeta_q^2 \rangle^{1/2} \sim H^2/(2\pi \dot{\phi}_0)$, we see the curvature perturbation associated with a hotspot differs primarily by $g/2$.
In this work we will choose $g\sim \mathcal{O}(1)$, so the two types of perturbations will be of the same order of magnitude.

\paragraph{CMB Anisotropy.}
After these fluctuation modes reenter the horizon, they source temperature anisotropies and give rise to localized spots on the CMB sky.
To compute the resulting anisotropies, we first write metric perturbations,
\begin{align}
ds^2 = -(1+2\Psi)dt^2+a^2(t)(1+2\Phi)\delta_{ij}dx^{i}dx^{j},    
\end{align}
in the Newtonian gauge.
The temperature fluctuations of the CMB corresponding to Fourier mode $\vec{k}$, pointing to direction $\hat{n}$ in the sky is given by,
\begin{align}
\Theta(\vec{k},\hat{n},\eta_0) = \sum_{l}i^l (2l+1)\mathcal{P}_l(\hat{k}\cdot\hat{n})\Theta_l(k,\eta_0). 
\end{align}
Here the multipole $\Theta_l(k,\eta_0)$ depends on the primordial perturbation $\zeta(\vec{k})$ and a transfer function $T_l(k)$ as,
\begin{align}
\Theta_l(k,\eta_0) = T_l(k)\zeta(\vec{k}), 
\end{align}
with $\eta_0$ denoting the conformal age of the Universe today.
Importantly, for our scenario $T_l(k)$ itself can be computed exactly as in the standard $\Lambda$CDM cosmology. It can be computed after taking into account the Sachs-Wolfe (SW), the Integrated Sachs-Wolfe (ISW), and the Doppler (Dopp) effect~\cite{dodelson2020modern},
\begin{equation}\label{eq:theta_l_master}
	\begin{aligned}		\Theta_l(k,\eta_0)& \simeq \left(\Theta_0(k,\eta_{\rm rec})+\Psi(k,\eta_{\rm rec})\right)j_l(k(\eta_0-\eta_{\rm rec}))\\
 & + \int_{0}^{\eta_0}d\eta e^{-\tau} \left(\Psi'(k,\eta)-\Phi'(k,\eta)\right)j_l(k(\eta_0-\eta))  \\
& + 3\Theta_1(k,\eta_{\rm rec})\left(j_{l-1}(k(\eta_0-\eta_{\rm rec}))-(l+1)\dfrac{j_{l}(k(\eta_0-\eta_{\rm rec}))}{k(\eta_0-\eta_{\rm rec})}\right)\\
& \equiv \left(f_{\rm SW}(k,l,\eta_0) + f_{\rm ISW}(k,l,\eta_0) + f_{\rm Dopp}(k,l,\eta_0)\right)\zeta(\vec{k})\,,
	\end{aligned}
\end{equation}
where $\tau$ is the optical depth. The above expression relates a primordial perturbations $\zeta$ to a temperature anisotropy $\Theta_l$.

\begin{figure}
    \centering
    \includegraphics[width=0.4\textwidth]{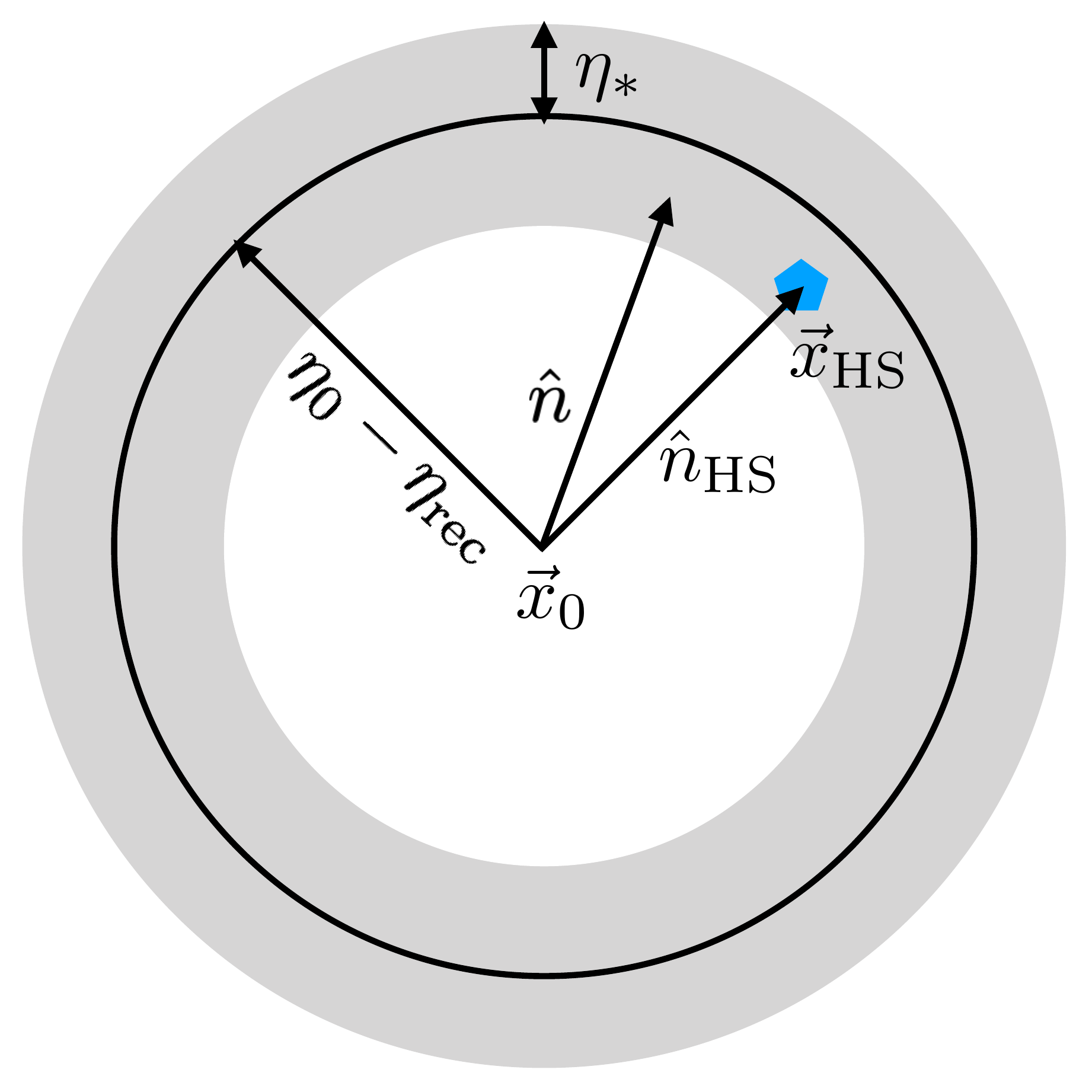}
    \caption{Representation of a hotspot on the CMB sky. Our location and the location of a hotspot are denoted as $\vec{x}_0$ and $\vec{x}_{\rm HS}$, respectively, defined with respect to an arbitrary coordinate system.
    The black circle denotes the surface of last scattering, located at  $\eta_{\rm rec} \approx 280$~Mpc in conformal coordinates.
    Due to momentum conservation, heavy particles are produced in pairs, and the distance between the two members of a pair can vary between 0 and $\eta_*$.
    Therefore, in our analysis we allow the two members to be anywhere within the gray shaded region.
    We compute the temperature profile of a hotspot as a function of direction of observation $\hat{n}$, with the hotspot center in the direction of $\hat{n}_{\rm HS}$.
    }
    \label{fig:hotspot}
\end{figure}

\paragraph{Temperature Anisotropy due to Heavy Particles.}
Regardless of the origin of $\zeta(\vec{k})$ is, we can compute $f_{\rm SW}(k,l,\eta_0)$, $f_{\rm ISW}(k,l,\eta_0)$, and $f_{\rm Dopp}(k,l,\eta_0)$ as in the standard $\Lambda$CDM cosmology. 
Thus converting the position space profile in Eq.~\eqref{eq:hs_simple} to momentum space and using Eq.~\eqref{eq:theta_l_master}, we can get the observed profile of a hotspot on the CMB sky.
This Fourier transform of the profile~\eqref{eq:hs_simple} can be written as,
\begin{align}
\langle\zeta_{\rm HS} (\vec{k}) \rangle = e^{-i\vec{k}\cdot\vec{x}_{\rm HS}}\frac{f(k\eta_*)}{k^3},
\end{align}
with a profile function
\begin{align}
f(x) = \frac{gH^2}{\dot{\phi}_0}(\text{Si}(x)-\sin(x)), \quad  \operatorname{Si}(x)=\int_0^x d t \sin (t) / t.
\end{align}
We parametrize the distance to the hotspot as,
\begin{align}
\vec{x}_0-\vec{x}_{\rm HS} = -(\eta_0 - \eta_{\rm HS})\hat{n}_{\rm HS}.   
\end{align}
Here $\vec{x}_0$ and $\vec{x}_{\rm HS}$ parametrize our and the hotspot locations, respectively, and  $\hat{n}_{\rm HS}$ points to the direction of the hotspot. 
The quantity $\eta_{\rm HS}$ denotes the location of the hotspot in conformal time with $\eta_0$ being the size of the present epoch.
In the earlier paper, we took the hotspot to be on the CMB surface and hence set $\eta_{\rm HS} = \eta_{\rm rec} \approx 280 ~$Mpc. In this work, we allow the hotspots to be away from the last scattering surface with $\eta_{\rm HS}$ between $\eta_{\rm rec}-\eta_*$ and $\eta_{\rm rec}+\eta_*$, and study their signals on the CMB surface. 
This set up is summarized in Fig.~\ref{fig:hotspot}.

As derived earlier, the temperature due to the hotspot is given by (dropping $\eta_0$ from the argument),
\begin{align}
\Theta(\vec{x}_0,\hat{n})=\int \frac{d^3\vec{k}}{(2\pi)^3}e^{i\vec{k}\cdot(\vec{x}_0-\vec{x}_{\rm HS})}\sum_{l}i^l (2l+1) \mathcal{P}_l(\hat{k}\cdot\hat{n})\left(f_{\rm SW}(k,l)+f_{\rm ISW}(k,l)+f_{\rm Dopp}(k,l)\right)\frac{f(k\eta_*)}{k^3}. 
\end{align}
Here $\hat{n}$ denotes the direction of observation.
The functions $f_{\rm SW}(k,l)$ and $f_{\rm ISW}(k,l)$ are extracted from the transfer function using {\tt CLASS}~\cite{Lesgourgues:2011re, Blas:2011rf} as in Ref.~\cite{Kim:2021ida}. Using the plane wave expansion,
\begin{align}
e^{-i\vec{k}\cdot\vec{r}}=\sum_{\ell=0}^\infty (-i)^l (2l+1) j_l(kr) \mathcal{P}_l(\hat{k}\cdot\hat{r}), \end{align}
and the relation
\begin{align}
\mathcal{P}_l(\hat{k}\cdot\hat{n})=\frac{4\pi}{(2l+1)}\sum_{m=-l}^{l}Y_{lm}(\hat{n})Y_{lm}^*(\hat{k}),    
\end{align}
we get:
\begin{eqnarray}\label{eq.thprofile}
\Theta(\vec{x}_0,\hat{n},\eta_{\rm HS}) &=& \frac{1}{2\pi^2} \int_0^\infty \frac{dk}{k}\sum_{l}j_{l}(k(\eta_0-\eta_{\rm HS}))(2l+1)\mathcal{P}_l(\hat{n}\cdot\hat{n}_{\rm HS})T_{\rm sum}(k,l) f(k\eta_*),   
\\
T_{\rm sum}(k,l)&\equiv&f_{\rm SW}(k,l)+f_{\rm ISW}(k,l)+f_{\rm Dopp}(k,l)\,.
\end{eqnarray}
Note $\Theta(\vec{x}_0,\hat{n},\eta_{\rm HS})$ depends on $\eta_{\rm HS}$, the location of the hotspot  -- which need not be on the last scattering surface as mentioned above. Given the spherically symmetric profile of the hotspot, the Doppler contribution to $\Theta(\vec{x}_0,\hat{n},\eta_{\rm HS})$ is small, from now on we only include the SW and ISW corrections for our analysis.

\begin{figure}[t!]
    \centering
    \includegraphics[width=0.6\textwidth]{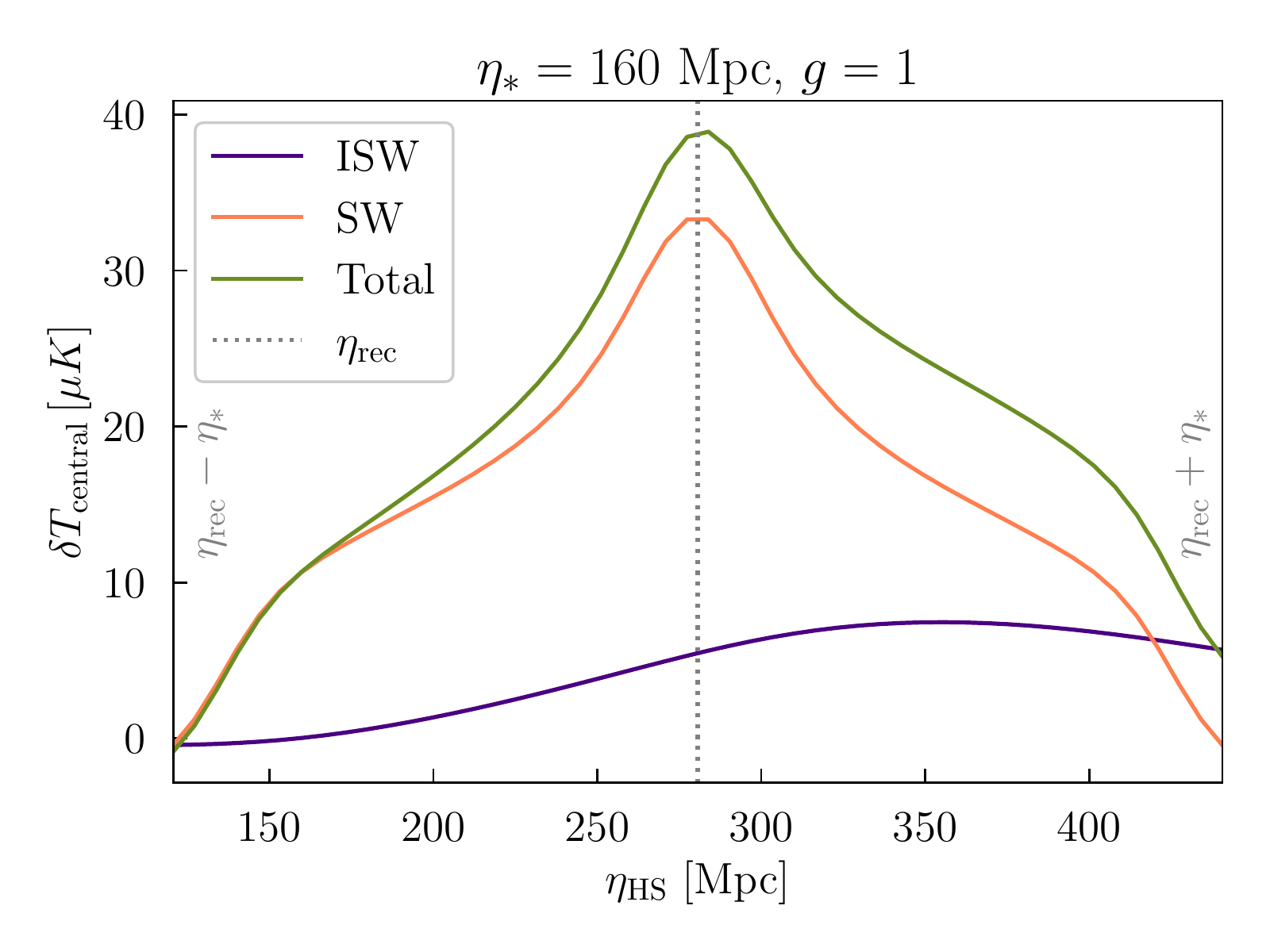}
    \caption{Central temperature $\Theta_{\rm central}\times T_{0}$ of a hotspot as a function of the (radial) location of the hotspot. We choose $\eta_*=160$~Mpc and $g=1$. The dotted gray line indicates the location of the recombination surface. Larger (smaller) $\eta_{\rm HS}$ implies the hotspots are closer to (further from) us. We also show contribution of the Sachs-Wolfe term (orange) and the Integrated Sachs-Wolfe term (purple) in determining the total temperature (olive). The left and right edges of the plot are at $\eta_{\rm HS} = \eta_{\rm rec} - \eta_*$ and $\eta_{\rm HS} = \eta_{\rm rec} + \eta_*$, respectively.
    }
    \label{fig:T_central}
\end{figure}

\begin{figure}[h]
    \centering
    \includegraphics[width=\textwidth]{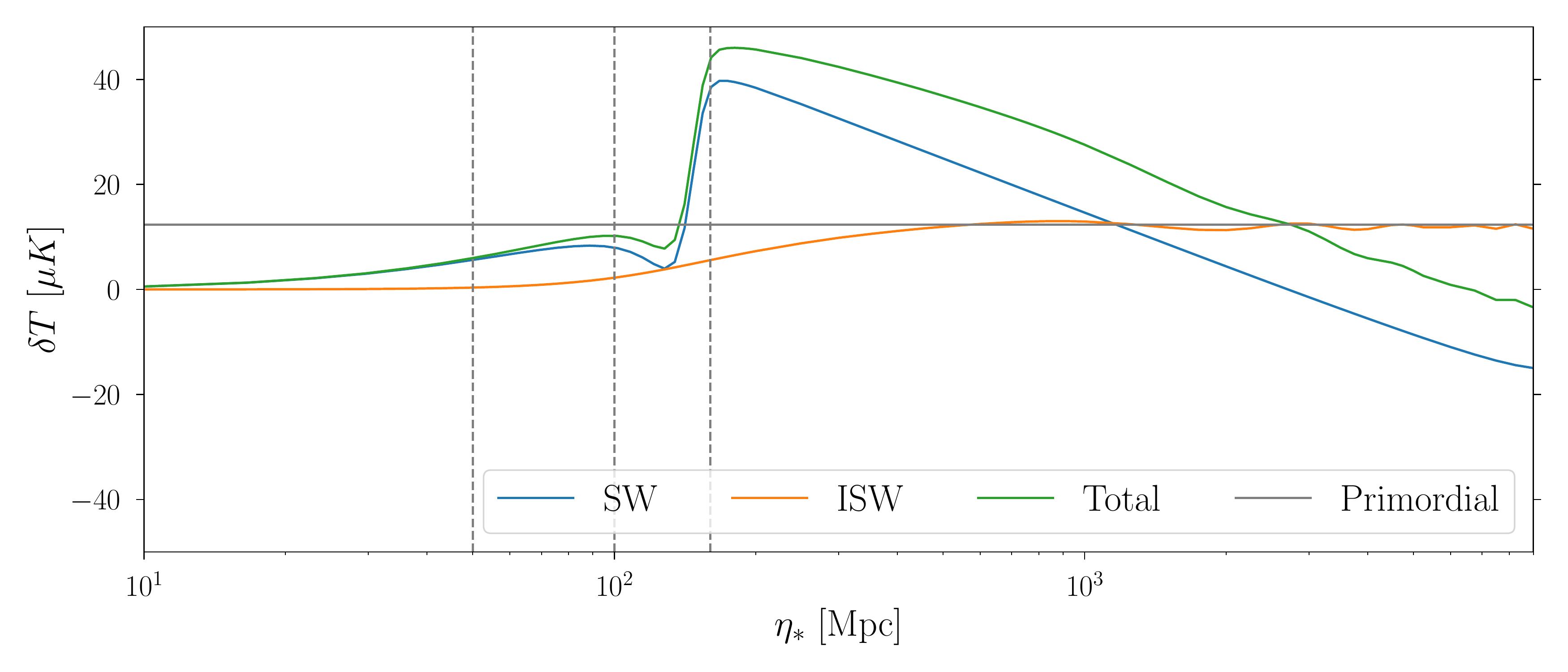}
    \caption{Central temperature (green) of a hotspot originating from a heavy particle for $g=1$, based on Eq.~\eqref{eq.thcentral} with $\eta_{\rm HS}=\eta_{\rm rec}$. 
    The green line illustrates the variation of the observed anisotropy as a function of the ``size'' of the hotspot, determined by the comoving horizon $\eta_*$ at the time of particle production. 
    The horizontal gray line gives a rough benchmark of the magnitude of the large-scale temperature anisotropy due to only the standard quantum fluctuations of the inflaton $(1/5)\langle \zeta_q^2\rangle$, without taking into account acoustic oscillations. 
    The dashed vertical gray lines show the benchmark choices for the hotspot size $\eta_*=50\,,100\,,160$~Mpc chosen in the subsequent discussion. We take the plot from Ref.~\cite{Kim:2021ida}.}
    \label{fig:transfer}
\end{figure}

\paragraph{Central Temperature.}
It is useful to compute the temperature anisotropy at the central part of a hotspot. 
To that end, we set $\hat{n}=\hat{n}_{\rm HS}$, implying $\mathcal{P}_l(\hat{n}\cdot\hat{n}_{\rm HS})= 1$, and
\begin{align}\label{eq.thcentral}
\Theta_{\rm central}(\vec{x}_0,\eta_{\rm HS}) = \frac{1}{2\pi^2} \int_0^\infty \frac{dk}{k}\sum_{l}j_{l}(k(\eta_0-\eta_{\rm HS}))(2l+1)T_{\rm sum}(k,l) f(k\eta_*).   
\end{align}
In Fig.~\ref{fig:T_central} we show the SW and ISW contributions to the central temperature as a function of $\eta_{\rm HS}$ after multiplying by the average CMB temperature $T_0=2.7$~K for $\eta_* = 160$~Mpc.
For completeness, we also show the central temperature in Fig.~\ref{fig:transfer}, as obtained in~\cite{Kim:2021ida}, as a function of hotspot size $\eta_*$, assuming the the hotspot is located on the  surface of last scattering. 
As we can see, the pair-produced CMB spots are indeed \textit{hot}spots when $\eta_*\lesssim$~Gpc. 
For $\eta_*>$~6600~Mpc \textit{cold}spots as opposed to \textit{hot}spots arise. This is because the negative SW contribution dominates the positive ISW contribution, with the combination being negative.

\begin{figure}[t!]
    \begin{center}
    \includegraphics[width=0.45\textwidth,clip]{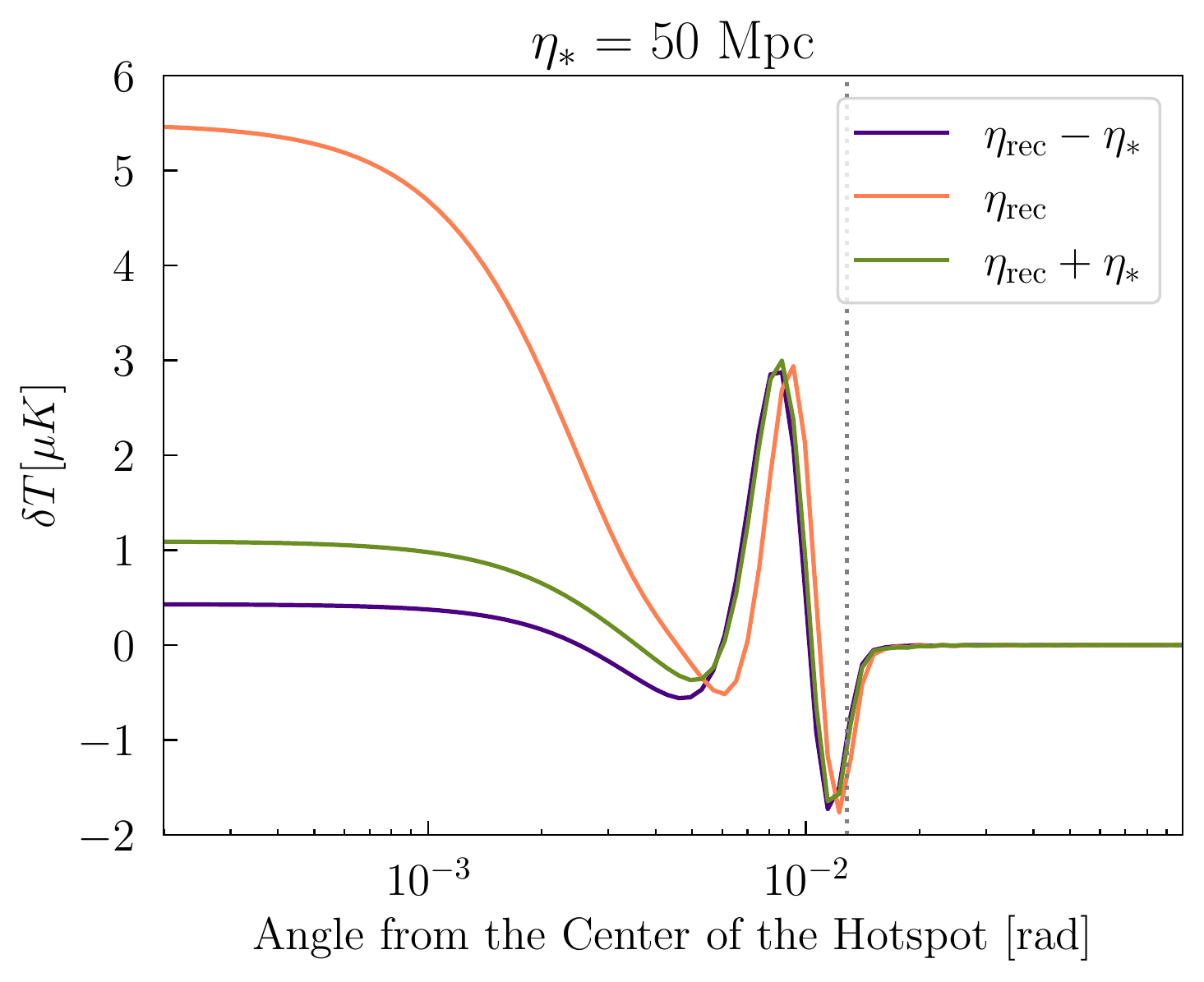}
    \includegraphics[width=0.45\textwidth,clip]{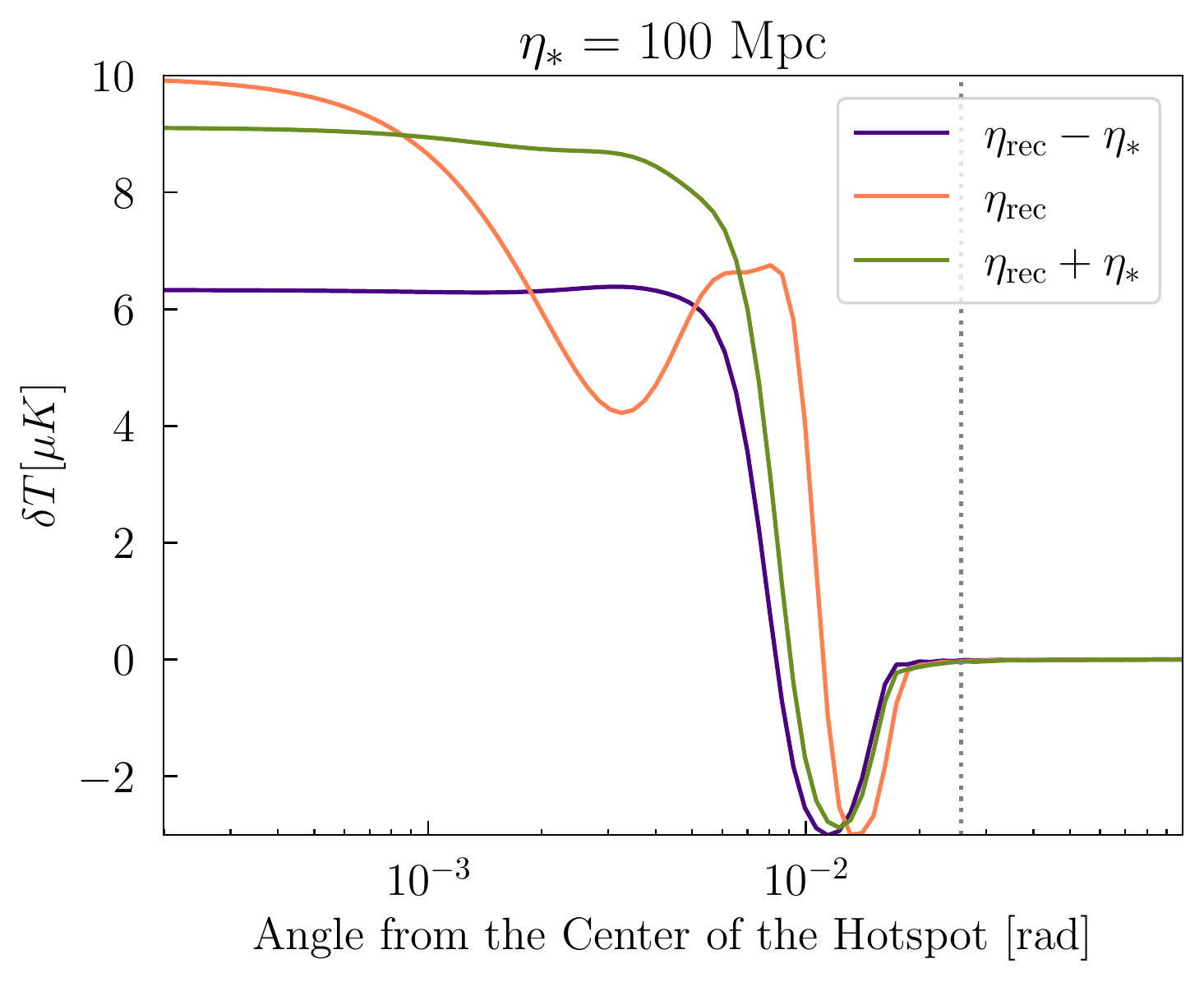}
    \includegraphics[width=0.45\textwidth,clip]{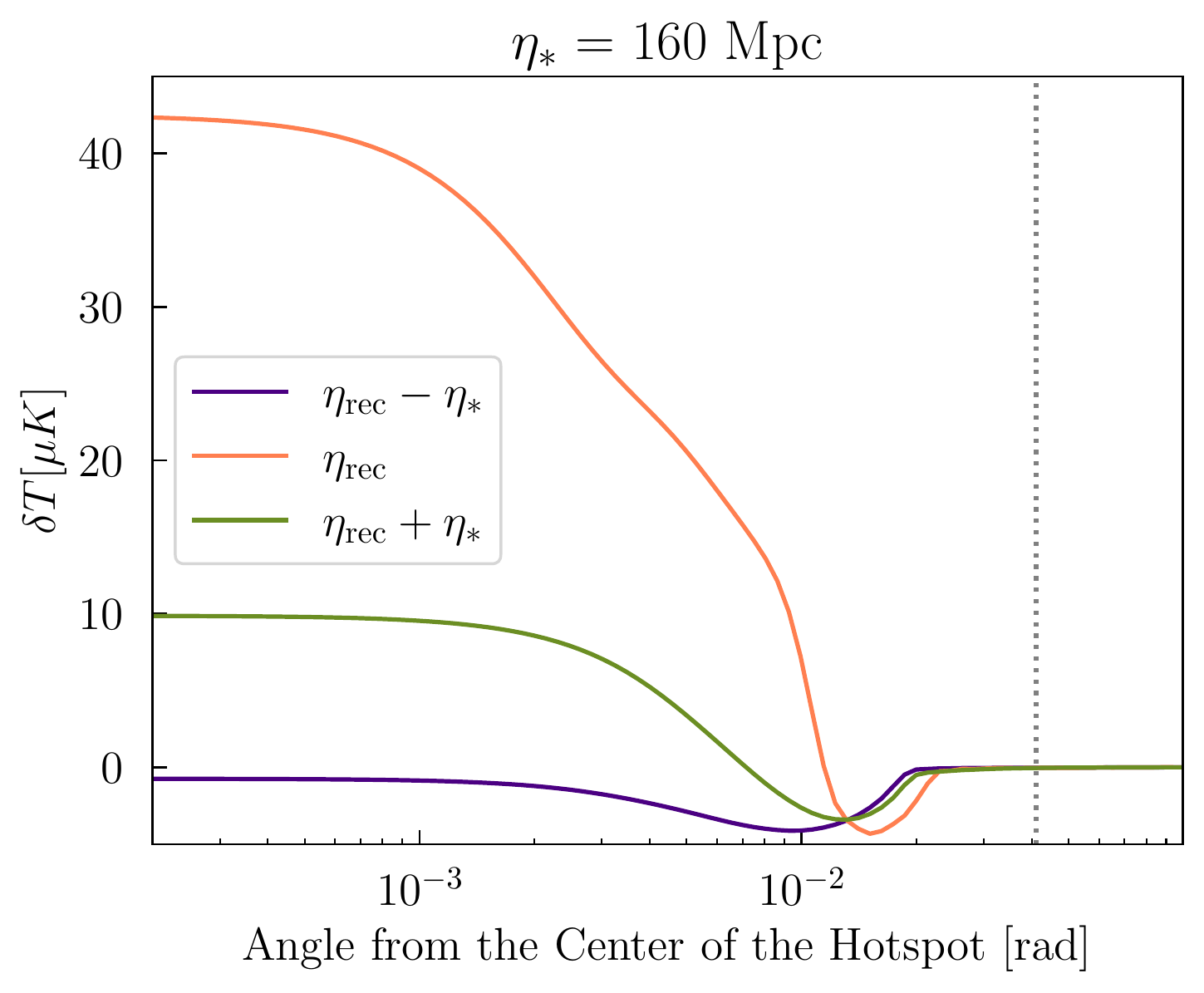}
    \end{center}
    \vspace{-0.1cm}
    \caption{\label{fig:delTvsrad} Radial profile of a single hotspot with the heavy particle position inside (olive), on (orange), and outside (purple) of the last scattering surface.
    The locations of these hotspots in conformal time are taken to be $\eta_{\rm rec} + \eta_*$, $\eta_{\rm rec}$, and $\eta_{\rm rec} - \eta_*$, respectively, as denoted by the labels.
    From upper left to bottom: horizon size for the hotspot production at $\eta_* = 50, 100, 160$~Mpc. 
    The plots assume the inflaton-$\chi$ coupling $g=1$.}
\end{figure}

\section{Simulation of the CMB and PHS Signals}\label{sec:image}
In order to design a PHS search, we simulate the PHS signal and CMB maps so that we can estimate the signal capture rate (`True Positive Rate'), and the background count for a CNN analysis.   
We notice that there are three types of backgrounds to consider for a PHS search: (i) the noise of the CMB detector,
(ii) the astrophysical foreground, and (iii) the background from the standard primordial fluctuations.

A realistic analysis needs to take into account detector noise and foregrounds. In our analysis, we consider profiles on relatively large angular scales, $\ell < 1000 $. For these scales current CMB temperature data, such as from Planck, is signal-dominated and we thus do not need to add instrumental noise to our simulations. The astrophysical foreground comes from compact objects such as galaxies, galaxy clusters, gas, and dust which can also produce localized signals. 
Part of these astrophysical foregrounds can be cleaned out due to their frequency dependence (for a review see, e.g., Ref.~\cite{Delabrouille:2007bq}). 
For the signal sizes that we consider, corresponding to $\ell < 1000 $, we do not expect significant astrophysical contamination after foreground cleaning and masking of the galactic plane, while for significantly smaller scales a detailed study of residual foregrounds and point sources would be required (see, e.g., Planck's component separation analysis~\cite{Planck:2013fzg}). 
In the following, we therefore only consider the background from the primordial, almost Gaussian, fluctuations when studying the PHS signal. 
This last type of background is `irreducible' in the sense that it will always be present, originating from the fluctuations of the inflaton itself.
We will assume the CMB maps are masked to reduce the astrophysical foregrounds and badly-conditioned pixels and retain only $60\%$ of the sky for the analysis. 
The number is similar to the sky fraction used in the Planck analysis~\cite{Aghanim:2018eyx}.

Unlike the analysis in~\cite{Kim:2021ida} that was based on a \Healpix~\cite{Gorski:2004by} simulation, in this work, we use the \Quicklens  package\footnote{\href{https://github.com/dhanson/quicklens}{https://github.com/dhanson/quicklens}} to simulate the CMB maps. 
\Quicklens allows us to work in the `flat sky approximation', neglecting sky curvature that is irrelevant to the size of the PHS profile we consider, as well as to draw sample maps with periodic boundary conditions to avoid complications due to masking.
\Quicklens can take a theoretical temperature power spectrum to produce mock flat sky CMB maps.
To provide an initial input, we use the \CLASS~(v3.2) package~\cite{Lesgourgues:2011re,Blas:2011rf} to compute a temperature anisotropy spectrum $C^{\rm TT}_{\ell}$ based on the Planck~2018~\cite{Planck:2018vyg} best fit $\Lambda$CDM parameters,
\begin{eqnarray} \label{eq.default}
\{\omega_{\rm cdm},\omega_b, h,  10^9A_s, n_s, \tau_{\rm reio}\}=\{0.120, 0.022, 0.678,  2.10, 0.966, 0.0543  \}\,.
\label{eq:params}
\end{eqnarray}
We will comment on the sensitivity of the CNN analysis to the $\Lambda$CDM parameters in Sec.~\ref{sec.training} and Appendix~\ref{app.LCDM}. 
We specify $\ell_{\rm max}=3500$ in the code for the maximum number of $\ell$-modes used for the image generation. As explained above, our signal profiles have support on length scales corresponding to an $\ell < 1000 $, where instrumental noise is negligible compared to the primary background from CMB and can thus be ignored. An application to significantly smaller angular scales would need to take into account the noise properties of the experiment.
We choose the image resolution such that 1 pixel $=10^{-3}$ radians to match Planck's angular resolution down to $\approx5$ arc minutes~\cite{Planck:2018nkj}. 
We also use the relation between the angle and the comoving length on the last scattering surface $\Delta\eta/\chi_{\rm rec}$.\footnote{In Ref.~\cite{Kim:2021ida}, the angular size of one pixel was obtained by matching the pixel number to the total degrees of freedom in the $\ell$-modes ($\ell_{\rm max}^2+\ell_{\rm max}=4\pi/\theta_{\rm pixel}^2$), together with the approximation $\ell_{\rm max}\simeq\eta_0/\eta_{\rm pixel}$. Although the matching reproduces the same angular resolution, the relation between $\ell_{\rm max}$ and $\eta_{\rm pixel}$ gives $\Delta\theta=\sqrt{4\pi}\Delta\eta/\chi_{\rm rec}$. Since $\ell_{\rm max}\simeq\eta_0/\eta_{\rm pixel}$ comes from the approximation of the $k$-mode integral with $j_{\ell}(k\,\chi_{\rm rec})$ and $k=2\pi/\eta$, the relation between the angle and length is less robust than $\Delta\theta=\Delta\eta/\chi_{\rm rec}$.}
For instance, if the separation between two hotspot centers is $160$~Mpc on the last scattering surface, the two centers are $12$ pixels away on the image, with $\chi_{\rm rec}=13871$~Mpc for Planck’s best-fit $\Lambda$CDM parameters. 

For the CNN analysis, we begin by generating $360^2$ pixel images, corresponding to a $ [-10.32^{\circ},10.32^{\circ}]$ region in longitude and latitude ($n_x=360$ in \Quicklens). 
We then cut out a $90^2$ patch from each of the $360^2$-sized maps. These non-periodic, smaller maps are then used for further analysis. In particular,
for our CNN analysis, we generate 160k training images, 40k validation $90^2$ pixel images, and an additional 5k test images to quantify the network performance. 
Training the neural network on smaller patches yields better training convergence and does not lead to loss of information as long as the characteristic size of the signal is smaller than the size of the patch.

\begin{figure}[t!]
    \begin{center}
    \includegraphics[width=0.3\textwidth,clip]{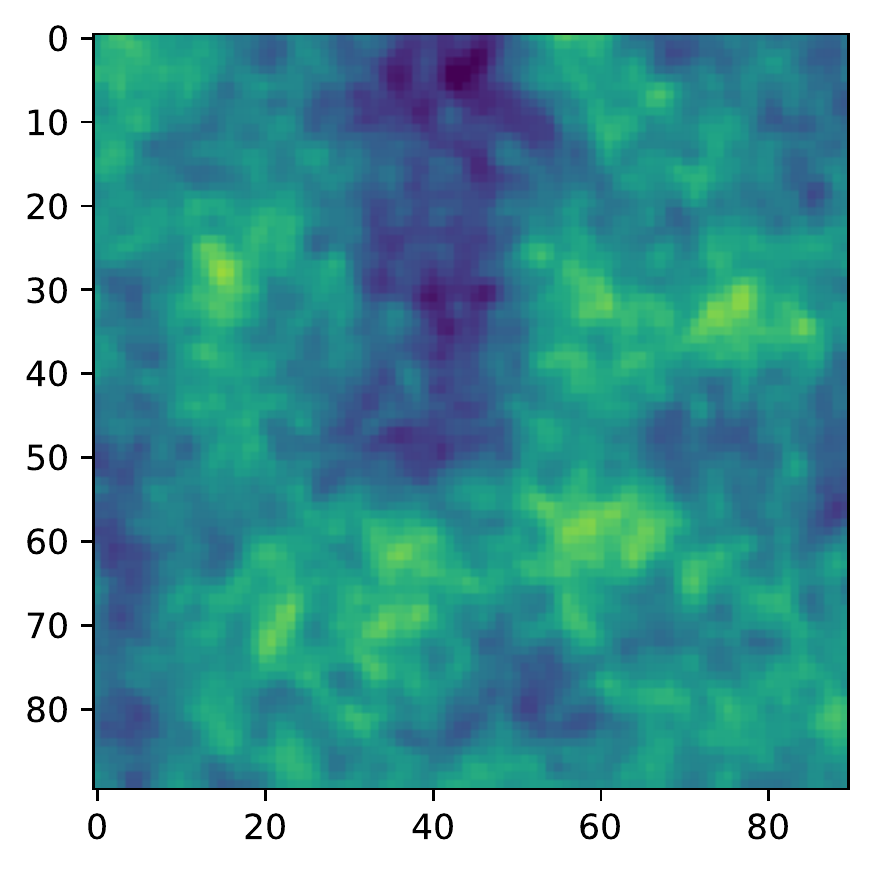}
    \includegraphics[width=0.3\textwidth,clip]{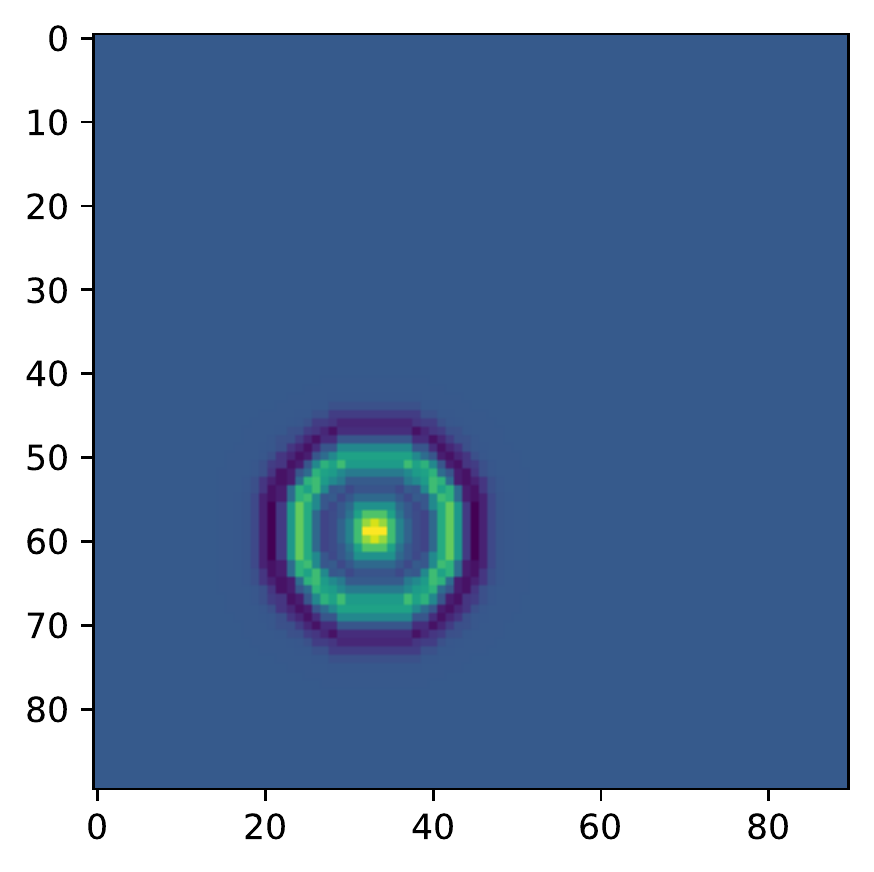}
    \includegraphics[width=0.3\textwidth,clip]{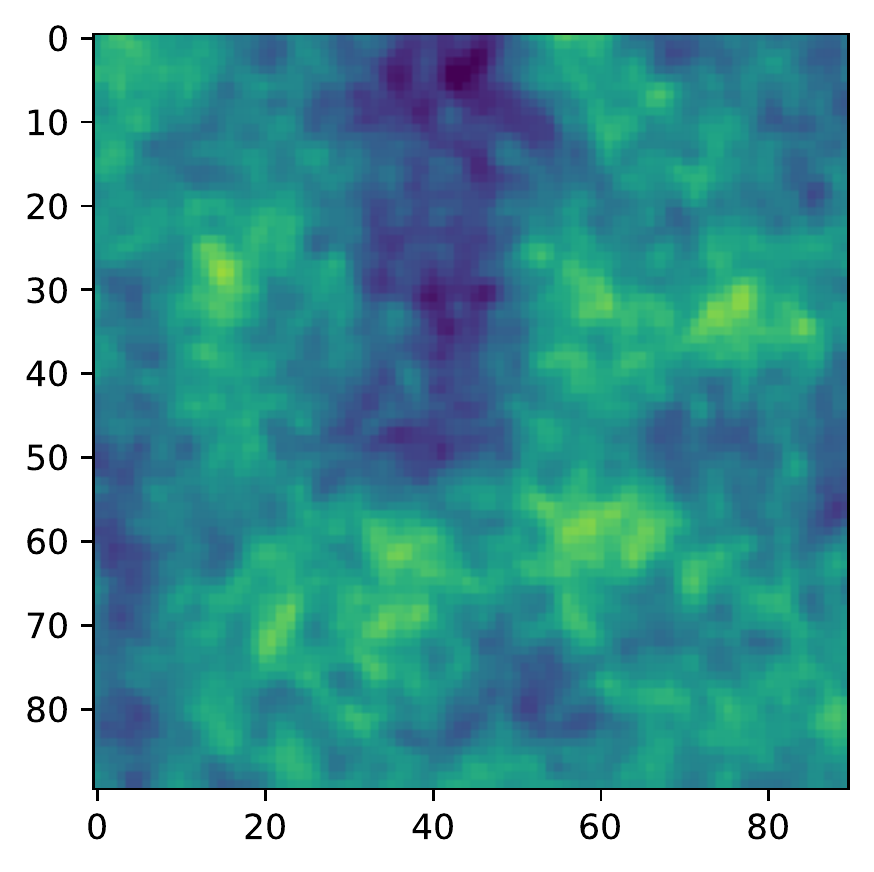}
    \\
    \includegraphics[width=0.3\textwidth,clip]{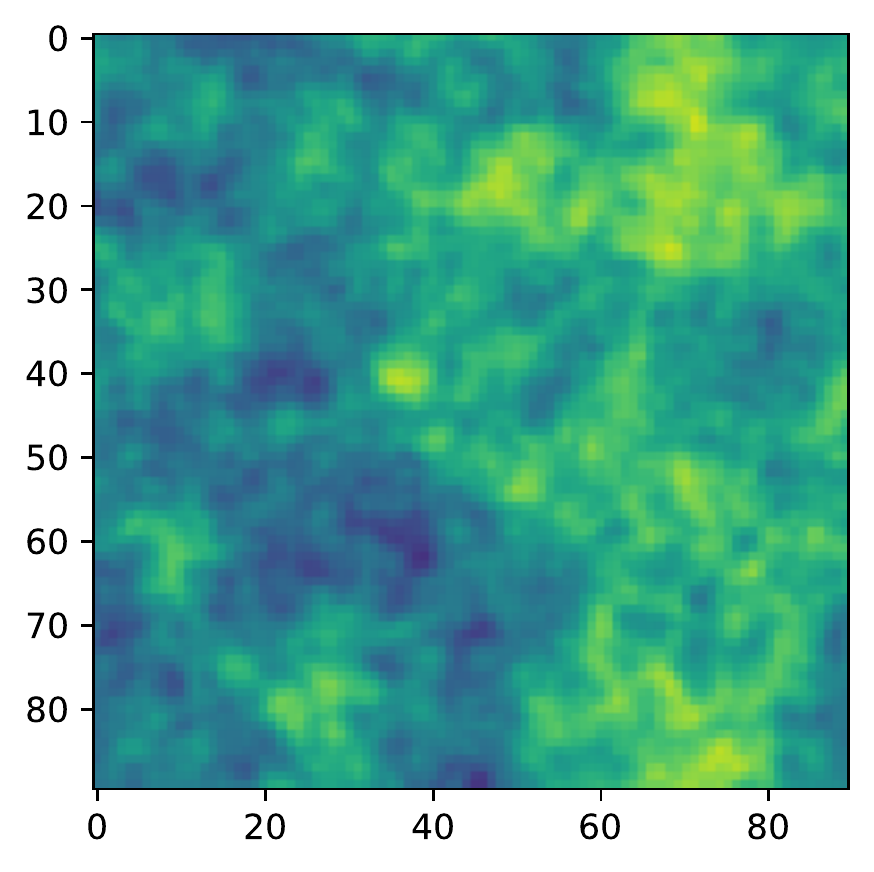}
    \includegraphics[width=0.3\textwidth,clip]{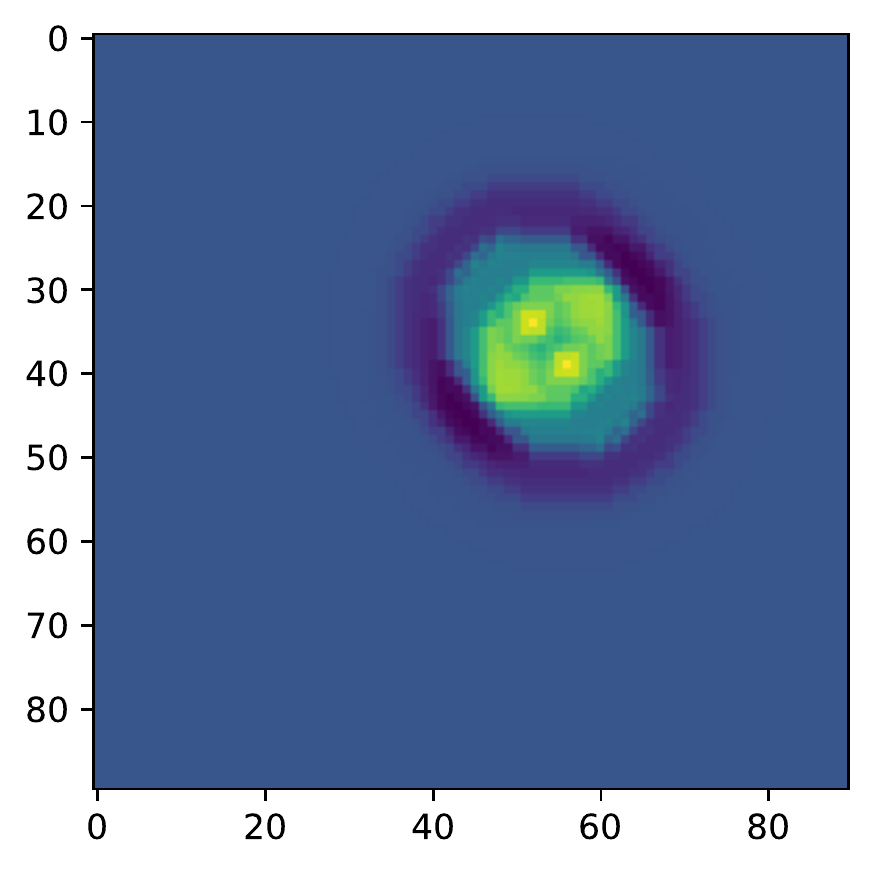}
    \includegraphics[width=0.3\textwidth,clip]{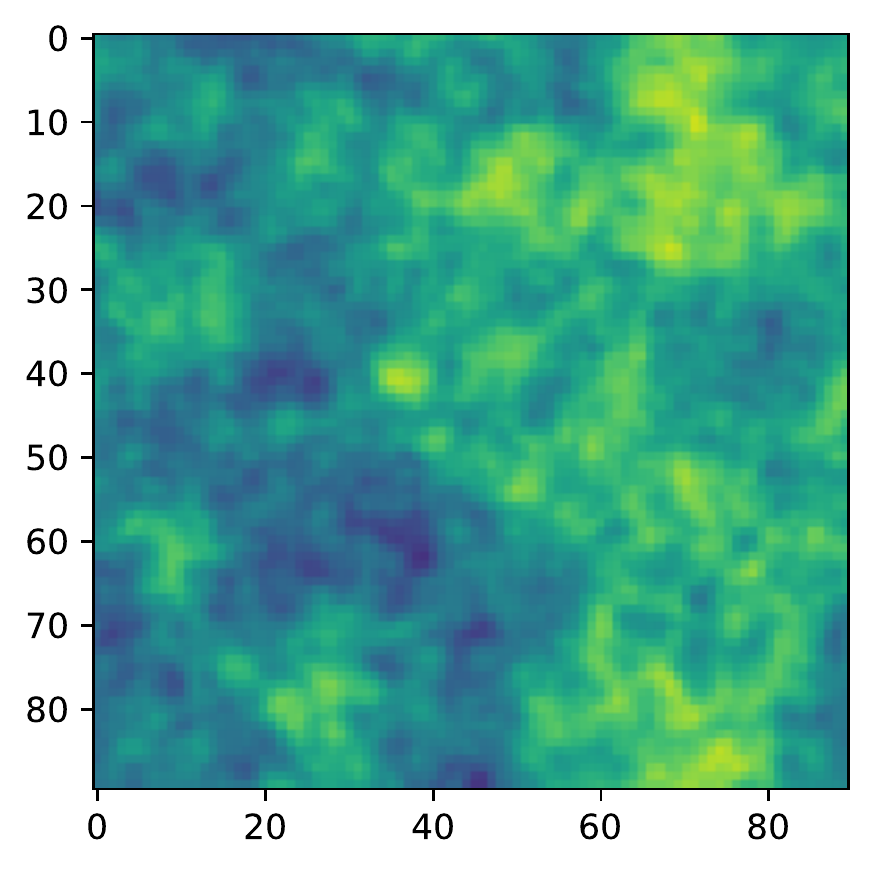}
    \\
    \includegraphics[width=0.3\textwidth,clip]{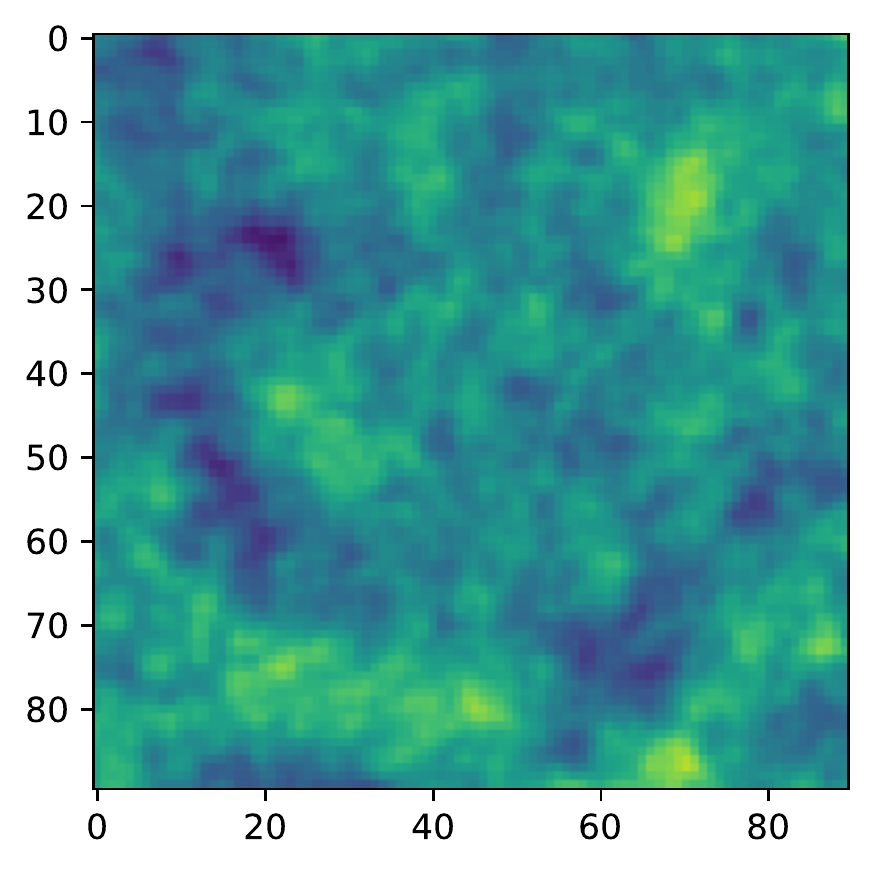}
    \includegraphics[width=0.3\textwidth,clip]{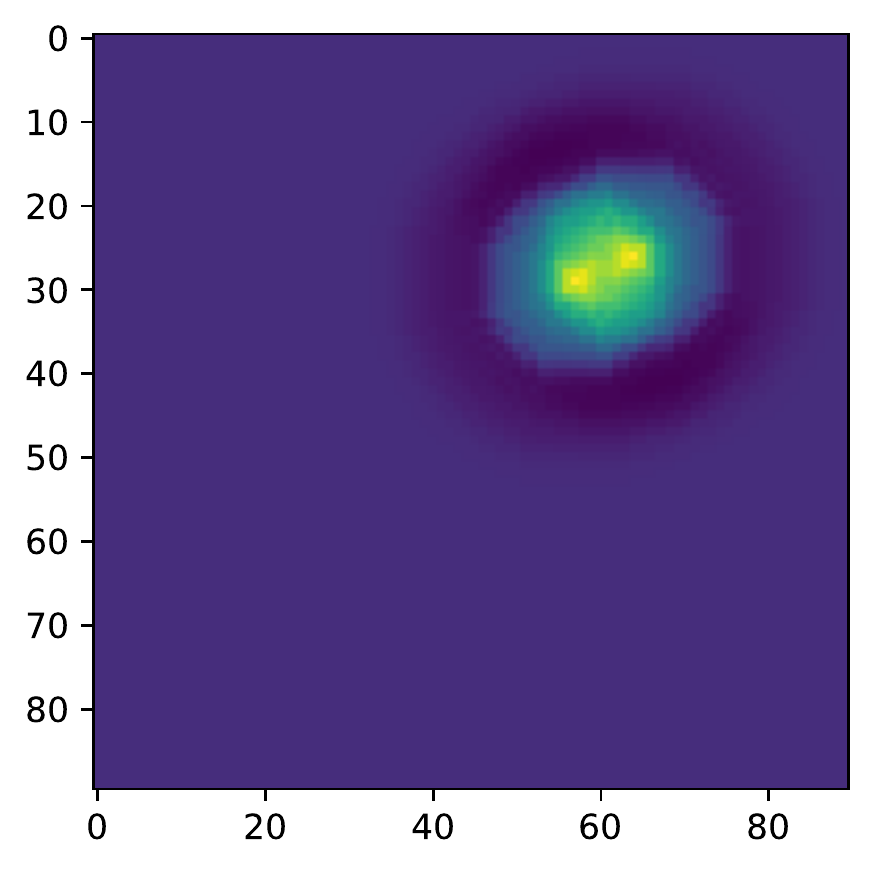}
    \includegraphics[width=0.3\textwidth,clip]{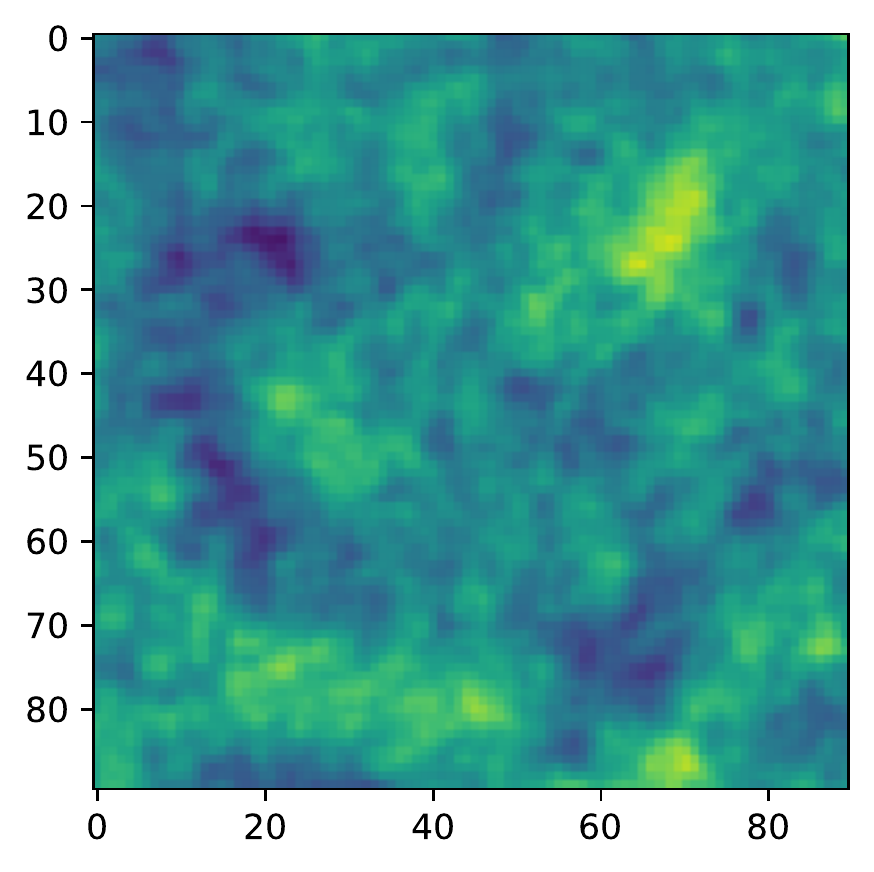}
    \end{center}
    \vspace{-0.1cm}
    \caption{\label{fig:PHSsample} Example plots of pure background from \Quicklens simulation (left), pure signals (middle), and signals with $g=4$ on top of the simulated background (right). The scalar particles are produced at comoving horizon sizes $\eta_*=50$~Mpc (top), $100$~Mpc (middle), $\eta_*=160$~Mpc (bottom). The signals at different benchmark $\eta_*$ have roughly the same size, as the $\eta_*$ dependence only enters logarithmically. The two hot spots are clearly separated for $\eta_* = 160\, \text{Mpc}$ and $\eta_* = 100\, \text{Mpc}$, while for $\eta_* = 50\, \text{Mpc}$ they overlap.}
\end{figure}

The profile of each of the PHS is described by Eq.~(\ref{eq.thprofile}), where the function depends on the distance to the hotspots $(\eta_0-\eta_{\rm HS})$ and the angle $\cos^{-1}(\hat{n}\cdot\hat{n}_{\rm HS})$, as defined in Fig.~\ref{fig:hotspot}. 
The overall magnitude of the signal temperature is proportional to the coupling $g$. 
When generating the signal, we require both the hotspots to be within a shell $\pm\eta_*$ around the last scattering surface as shown in Fig.~\ref{fig:hotspot}. 
For example, when studying the case with $\eta_*=160$~Mpc, we first divide the $\pm 160$~Mpc region into 50 concentric annuli, each having equal thickness.
We then choose the first hotspot from a pair to lie on any of these 50 annuli with equal probability.
The second member is then chosen anywhere within a sphere of radius $\eta_*$ centered on the first hotspot, again with a uniform random distribution.\footnote{Our motivation for the uniform distribution is driven by Eq.~\eqref{eq.nintgl}. There, the integral is dominated by $k_p \sim M_0$, as small $k_p$ are suppressed by the $k^2_p$ factor and large $k_p$ are exponentially squashed. Assuming that $\chi$ are produced semi-relativistically and can travel a horizon scale distance, we expect a uniform distribution of hotspots within $\pm \eta_*$ of the last scattering surface.}
A pair is kept for further analysis only if both the spots of the pair falls within the $\pm\eta_*$ shell of the last scattering surface.
Since the distribution in a 3D volume allows hotspots to orient along the line-of-sight direction, the average separation between the two hotspots projected on the last scattering surface is smaller than the separation assumed in Ref.~\cite{Kim:2021ida} that only considered PHS on the last scattering surface. 

Once we generate PHS images with random orientation and separation between two hotspots, we pixelate them and add the PHS image to the simulated CMB maps to produce the signal image. 
We follow this procedure for all the signal images in our study.
In this work, we study benchmark models with horizon sizes
\beq
\eta_*=50,\,100,\,160~{\rm Mpc}\,,
\eeq
and couplings from $g=1$ to $4$. Specifying $g$ and $\eta_*$ sets the overall temperature and the profile of the hotspot, a la Eq.~\eqref{eq.thprofile}. 
Within the approximations we've made in Sec.~\ref{sec.PHS}, the remaining model parameter, $M_0$, only affects the overall number of hotspots $N_{\rm PHS}$ (through Eq.~\eqref{eq.Nspots}). 
Going forward, we will compute the number of hotspots that can be hidden within the background fluctuations for given benchmark coupling and $\eta_*$. Then, using Eq.~\eqref{eq.Nspots}, the upper bounds on $N_{\rm PHS}$ can be translated into lower bounds on $M_0$. 
As an illustration of what a benchmark PHS looks like, in Fig.~\ref{fig:PHSsample} we show examples of the CMB background (left), PHS signal (middle), and the signal plus background (right) for $g=4$ with different choices of $\eta_*$. Note that it is difficult to identify the signals by eye in the plots on the right, even with such a large coupling. 

Compared to Ref.~\cite{Kim:2021ida}, the benchmark $\eta_*$ values are identical, but we choose smaller values of the coupling $g$.
This is because we find the CNN analysis is much more powerful than the `cut and count' method adopted in Ref.~\cite{Kim:2021ida}, and therefore capable of identifying fainter hotspots. 
We chose the benchmark $\eta_*$ values to test out a variety of different PHS; $\eta_*=160\,\text{Mpc}$ hotspots have a very high central temperature (Fig.~\ref{fig:transfer}), while $\eta_*=50\,\text{Mpc}$ hotspots are significantly cooler and have smaller inter-spot separation. 
The choice  $\eta_*=100\,\text{Mpc}$ sits between these for comparison.

\section{Identifying Pairwise Hotspots with CNN}\label{sec.CNN}
\begin{figure}[t!]
    \begin{center}
    \includegraphics[trim={100 250 400 170},clip,width=\textwidth]{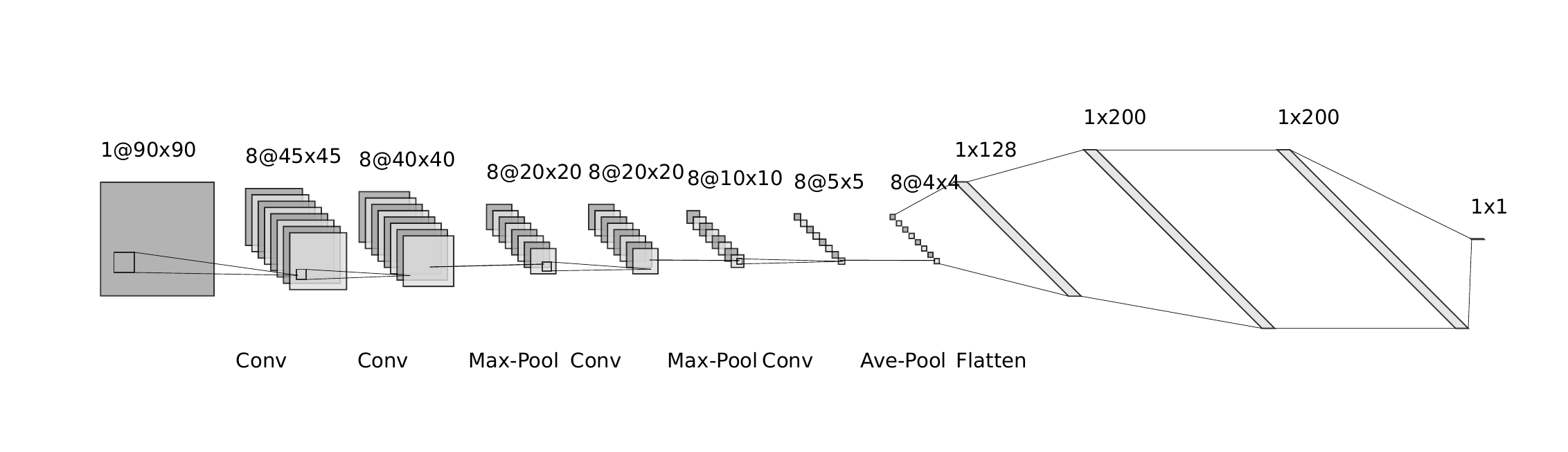}
    \end{center}
    \vspace{-0.5cm}
    \caption{A schematic architecture of the CNN used in this work. We applied two convolutional layers in series; first, 8 kernels with size of $16\times 16$ and stride of $2\time 2$ are applied, then, 8 independent kernels size of $8\times 8$ yields feature map of $40\time 40\time 8$. Next, we apply a max-pooling using the kernel and stride size of $ 2\times 2$, which subsequently reduces the image dimension down to $20 \times 20\times 8$. Processed images get further reduced by going through 2D convolution and max-pooling, further reducing the size of the image to $5\times 5\times 8$. After 4 sets of total convolution followed by average pooling, the final feature maps are flattened to feed into fully connected network form, and the final network ends with single output value which sits between 0 and 1. Throughout the network, we use the rectified linear unit (ReLU) function~\cite{pmlr-v15-glorot11a} to introduce non-linearity, except for the output layer which has a sigmoid activation function suitable for the binary classification.}
\label{fig:CNN_architecture}
\end{figure}
\begin{figure}[t!]
    \begin{center}
    \includegraphics[width=1.0\textwidth,clip]{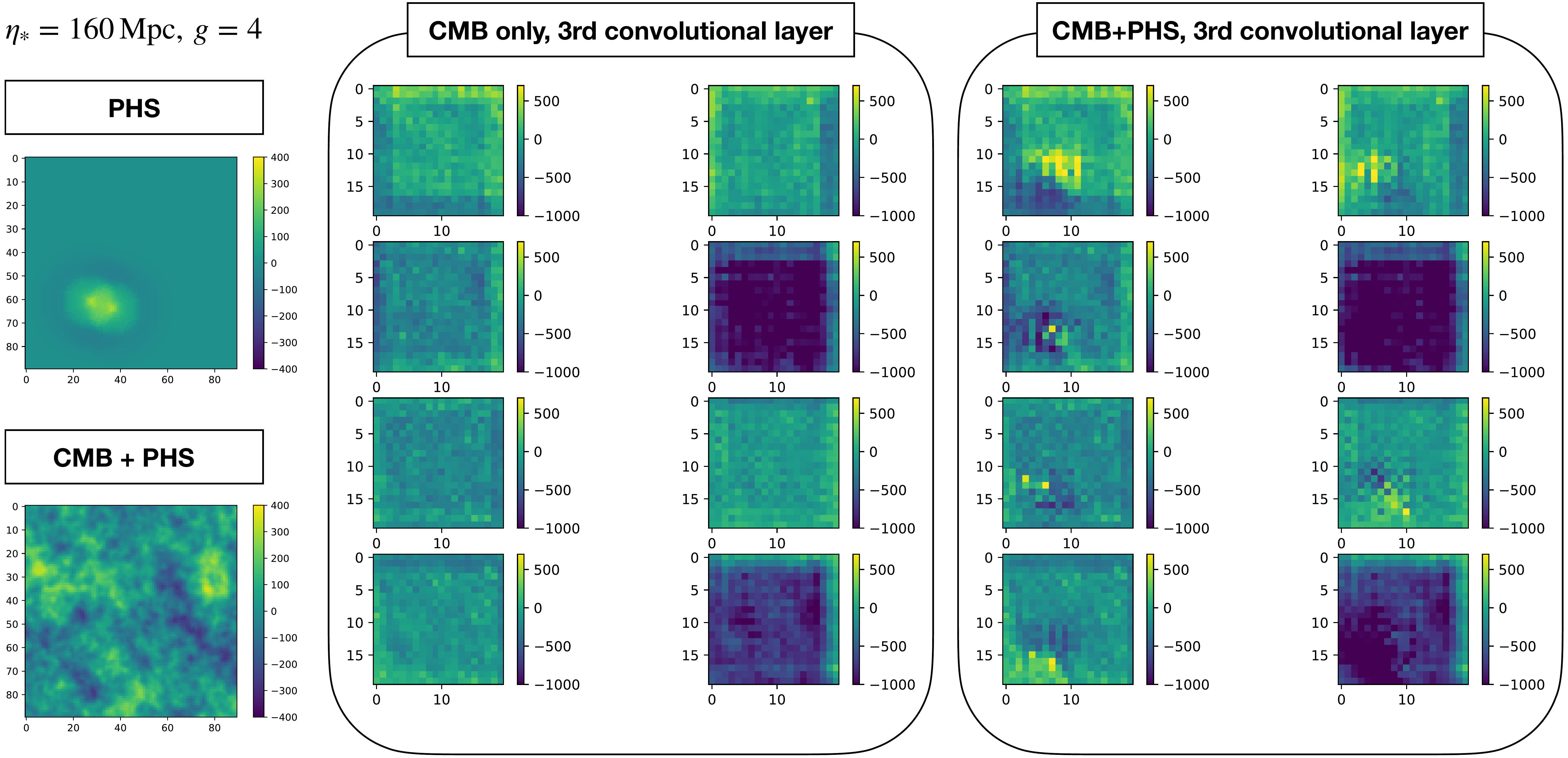}
    \end{center}
    \vspace{-0.5cm}
    \caption{\label{fig:truevscnn50} Comparison between true and the CNN feature maps with and without implanted signals. The left plots show the PHS signal and signal plus the CMB background. The middle and right plots show feature maps after going through three convolutional layers. The enhanced signal locations on the feature maps on the right align with the true location of the hotspots after rescaling the pixel coordinates with respect to the relative size between the 3rd layer ($20^2$-pixels) and the original image ($90^2$-pixels). Here we take $\eta_* = 160$~Mpc, $g=4$, and $\eta_{\rm HS}=\eta_{\rm rec}$ for both the spots.}
\end{figure}

In this section we describe the training process for the CNN using $90^2$ pixel images, and discuss some qualitative properties of the training result. 
We then apply the trained network to a larger sky map and present results on the upper bound on the number of PHS for given values of $\eta_*$ and $g$.
We end the section with some comparisons between the CNN and a matched filter analysis.

\subsection{Network Training on Small Sky Patches}\label{sec.training}
CNNs are one of the most commonly used deep neural networks specialized for image recognition~\cite{LeCun:2012,Krizhevsky:2012}.
In this study, we build the network using PyTorch~\cite{NEURIPS2019_9015} with the structure shown in Fig.~\ref{fig:CNN_architecture}. 
The network takes a CMB or CMB+PHS image as an input and outputs a single value between 0 and 1, which can be interpreted as the probability of the input image containing the PHS. 
We train the network on 160k images (see Sec. \ref{sec:image}), half of which contain a single pairwise hotspot profile on top of the CMB and the rest are CMB-only images.
For optimization, we use a binary cross entropy loss function, commonly used for binary classification, along with Adam optimizer~\cite{Kingma:2014vow} and $10^{-4}$ learning rate.

We train the network using PHS signals with $g=3$ for all the three values of $\eta_*$ individually. One may wonder how well a network trained on one $g$ value will generalize to different values without retraining. As the CNN (unlike the matched filter discussed below) is nonlinear, extrapolation to values of $g$ other than what was used for training is not guaranteed to be optimal. On the other hand, training a CNN for each possible benchmark input is time- and resource-intense. Empirically, we find that the network trained at $g=3$ works well over a wide range of $g$ values, perhaps because the network learns to analyze the shape rather than the amplitude of the profile. 
In a fully optimal analysis one would want to retrain the neural network over a grid of $g$ values.

To get some idea for how the CNN discriminates between signal and background images, we show the feature maps from the first three convolutions in Fig.~\ref{fig:truevscnn50} for $\eta_* = 160\, \text{Mpc}$ and $g=4$. 
As we can see proceeding from left to right, the trained network does amplify the signal region compared to the background-only image, and the convolutional layers can emphasize the correct locations of each spot in the feature map.

To quantify the performance of the CNN, we generate a test sample of $5$k CMB-only maps and 5k CMB+PHS maps, each having $90^2$ pixels.
For a CMB+PHS map, we inject one randomly oriented and located PHS in the CMB map. 
The PHS signal occupies $\mathcal O(50^2)$ pixels in the examples that we study, and thus the $90^2$-pixels image is only slightly larger than the signal. 
When an image has network output $>0.5$, we count it as an identified signal map. 
We call the signal capture rate (True Positive Rate, $\epsilon_{S,90^2}$) as the fraction of CMB+PHS images being correctly identified as signal maps, and define the fake rate (False Positive Rate, $\epsilon_{B,90^2}$) as the fraction of CMB-only images being wrongly identified as signal maps,\footnote{In the actual search, there can be more than one PHS in a $90^2$-pixels region, and the CNN would still count the region to be one signal map. We verify that the signal capture rate would increase if there are more PHS in the image. When we study the sensitivity of the CNN search, having additional PHS around the same location will help the search, and this makes our analysis based on having one PHS in a $90^2$-pixels image to be conservative. Moreover, given that the CNN search can probe PHS with a small number of signals on the CMB sky, the probability of having additional PHS around the same location is small. Therefore, counting the number of $90^2$-pixels regions should give a good approximation of the PHS in the analysis.}
\begin{align}
\label{eq:90sqefficiency}
\epsilon_{S,90^2} &= \frac{\text{number of signal-injected images with CNN output $>$ 0.5 }}{\text{total number of signal-injected images}},\nonumber \\
\epsilon_{B,90^2} &= \frac{\text{number of background-only images with CNN output $>$ 0.5}}{\text{total number of background-only images}}.
\end{align}

\begin{figure}[t!]
    \begin{center}
    \includegraphics[width=0.47\textwidth,clip]{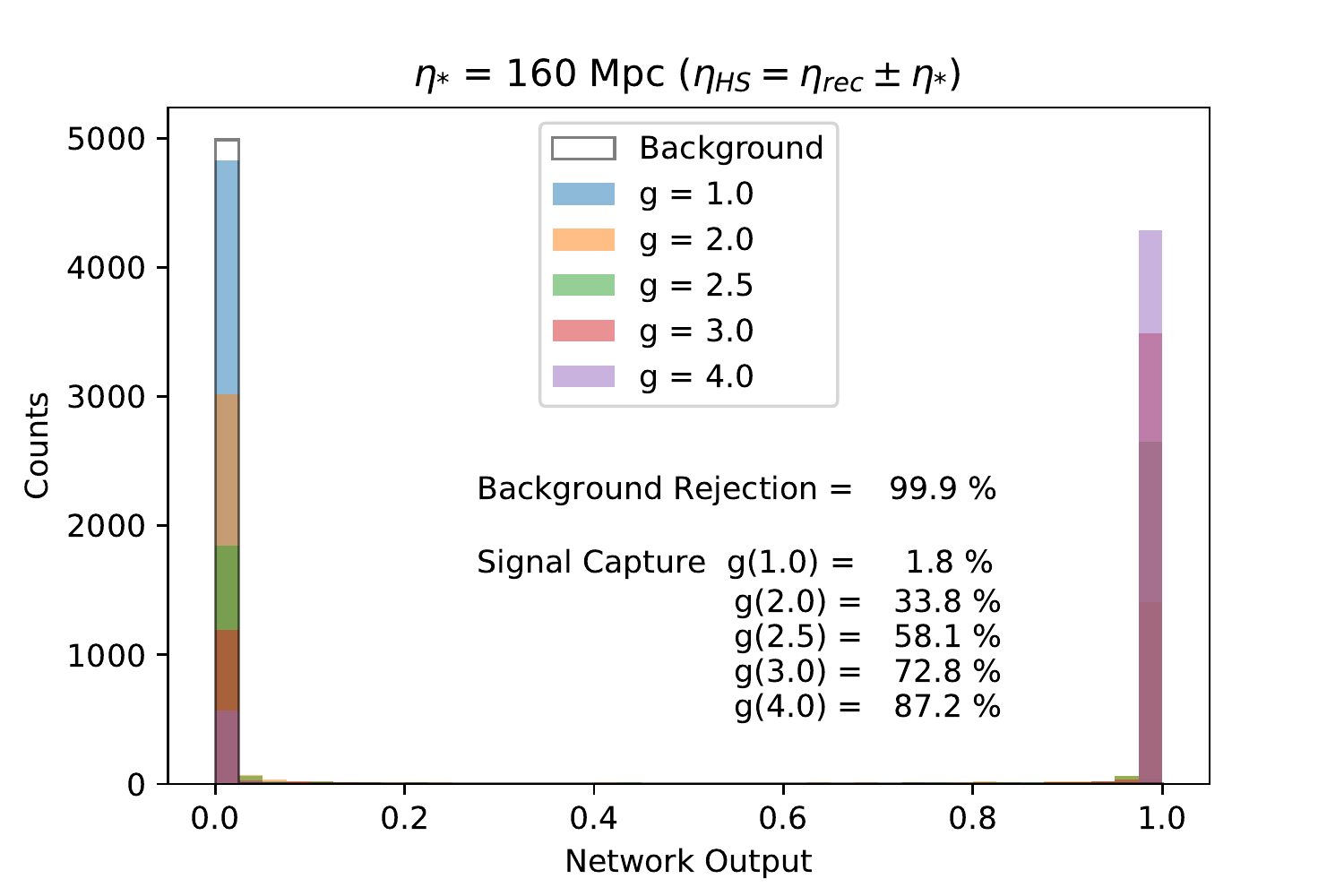} 
   \includegraphics[width=0.47\textwidth,clip]{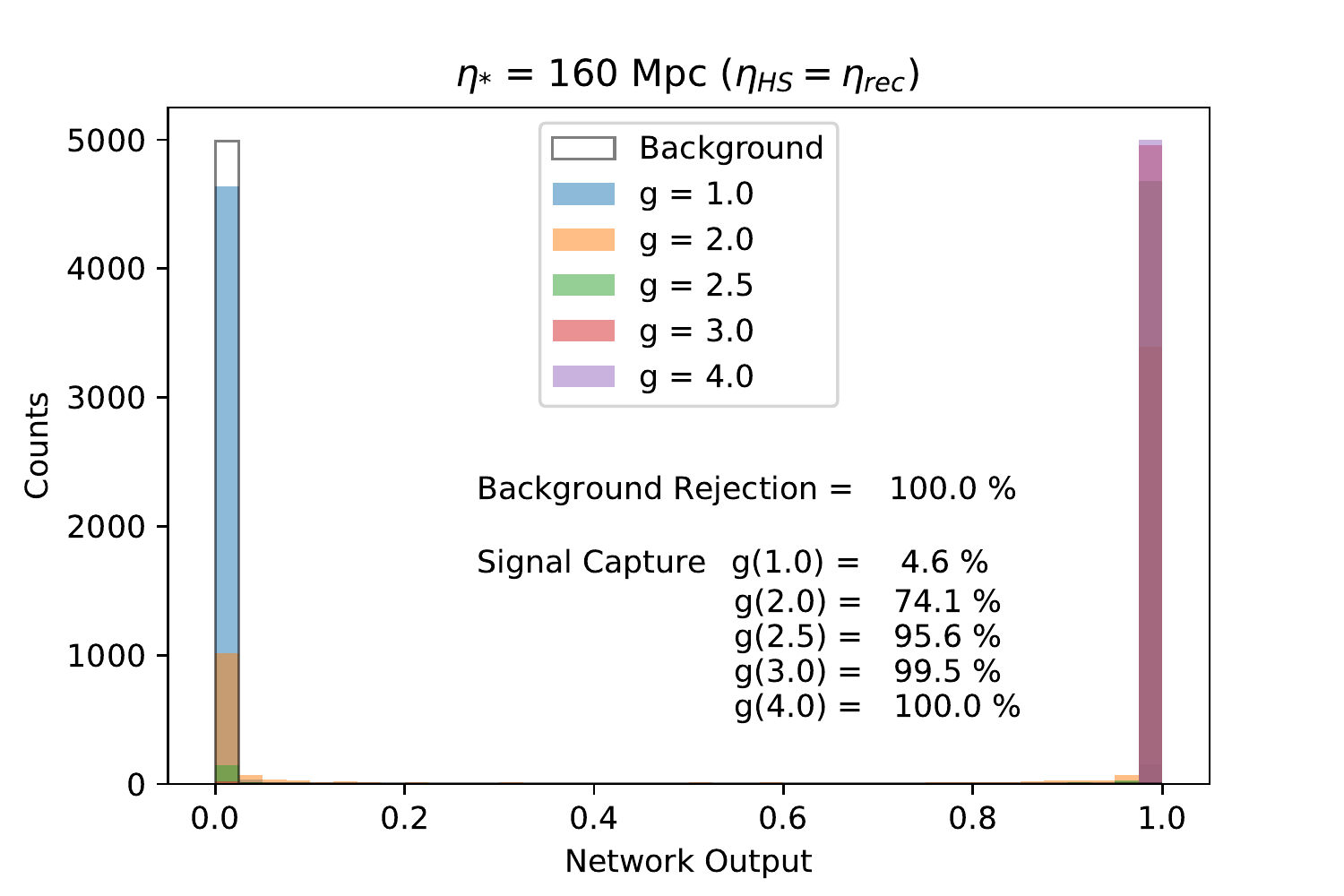} 
   \includegraphics[width=0.47\textwidth,clip]{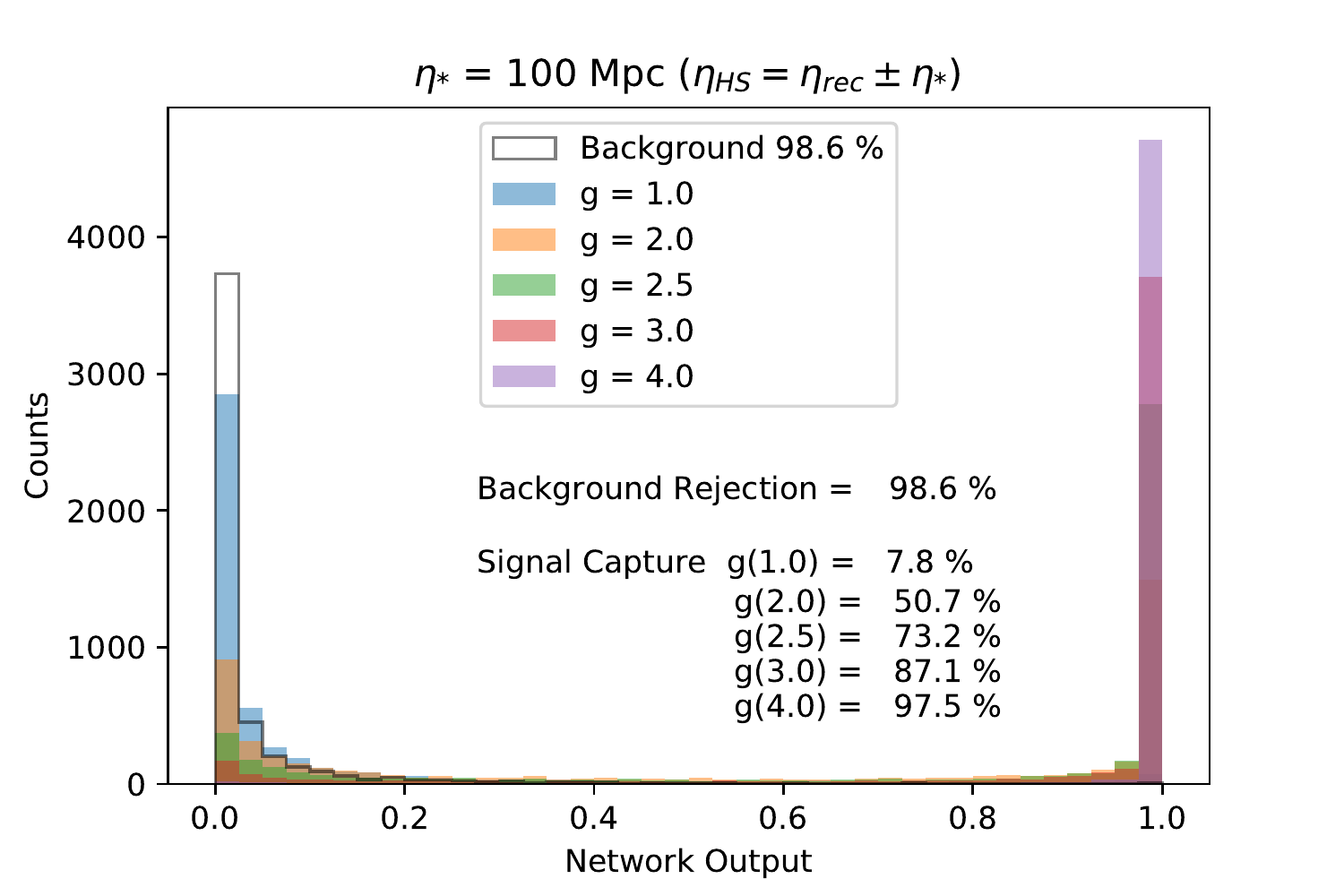} 
    \includegraphics[width=0.47\textwidth,clip]{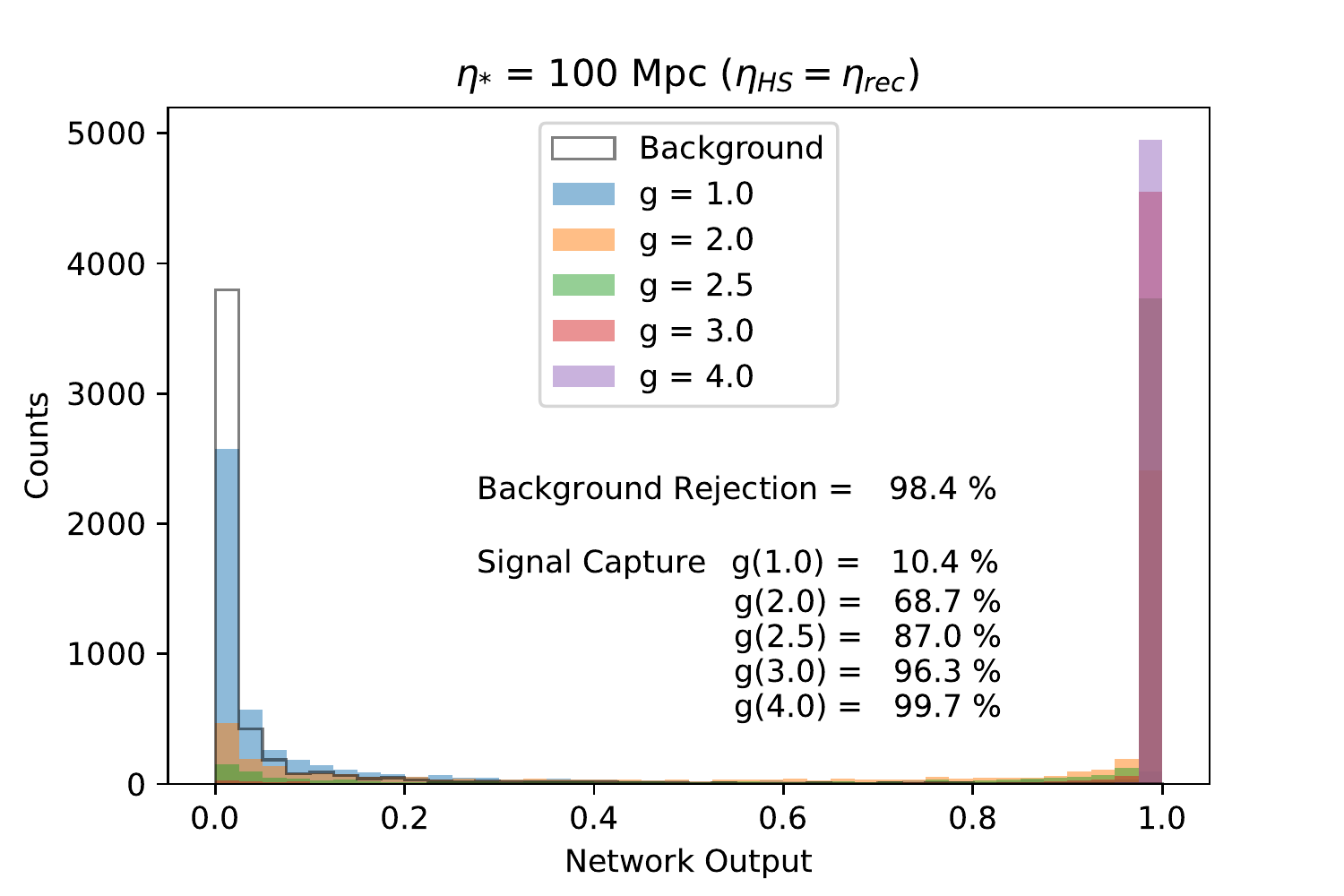} 
 \includegraphics[width=0.47\textwidth,clip]{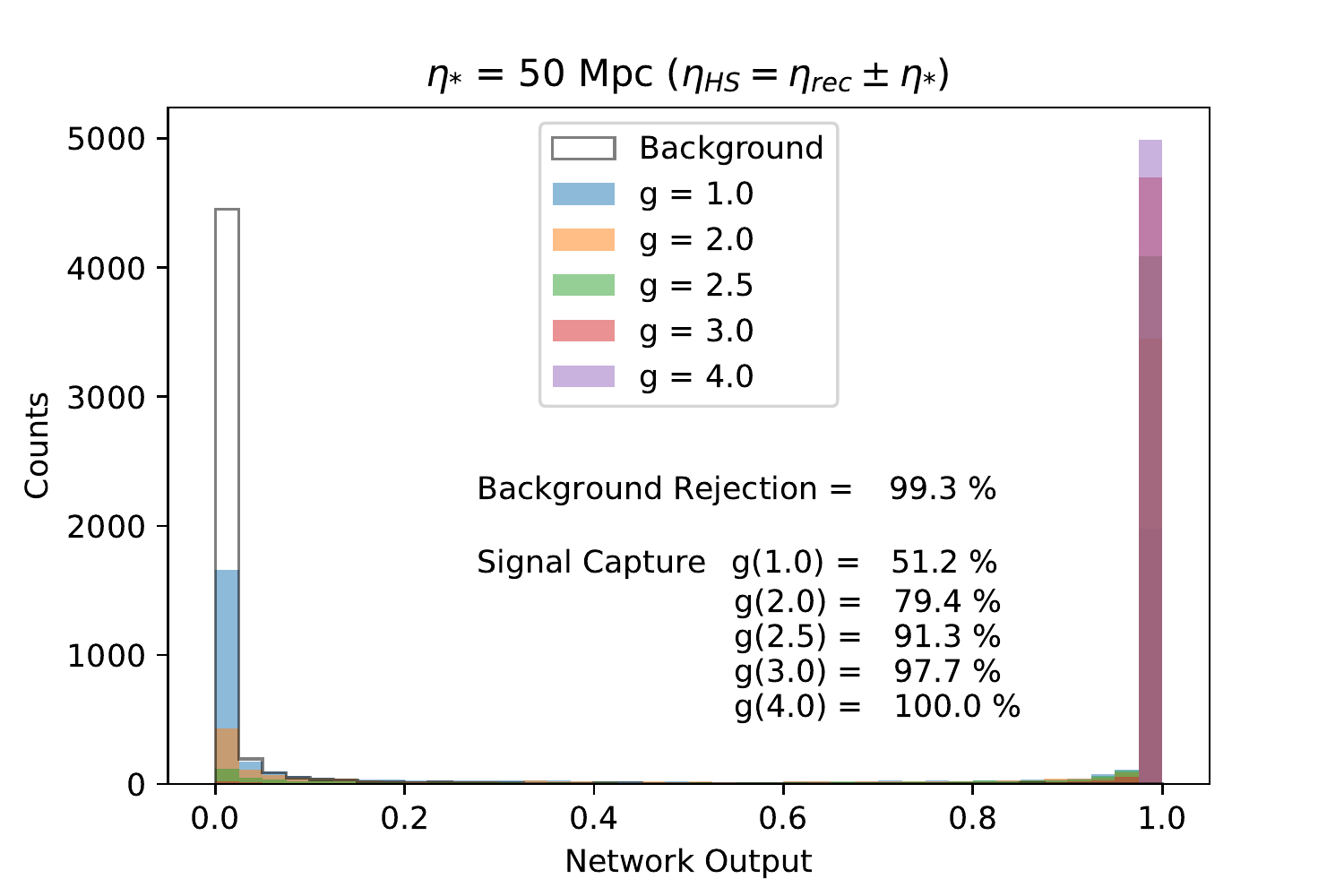}     
    \includegraphics[width=0.47\textwidth,clip]{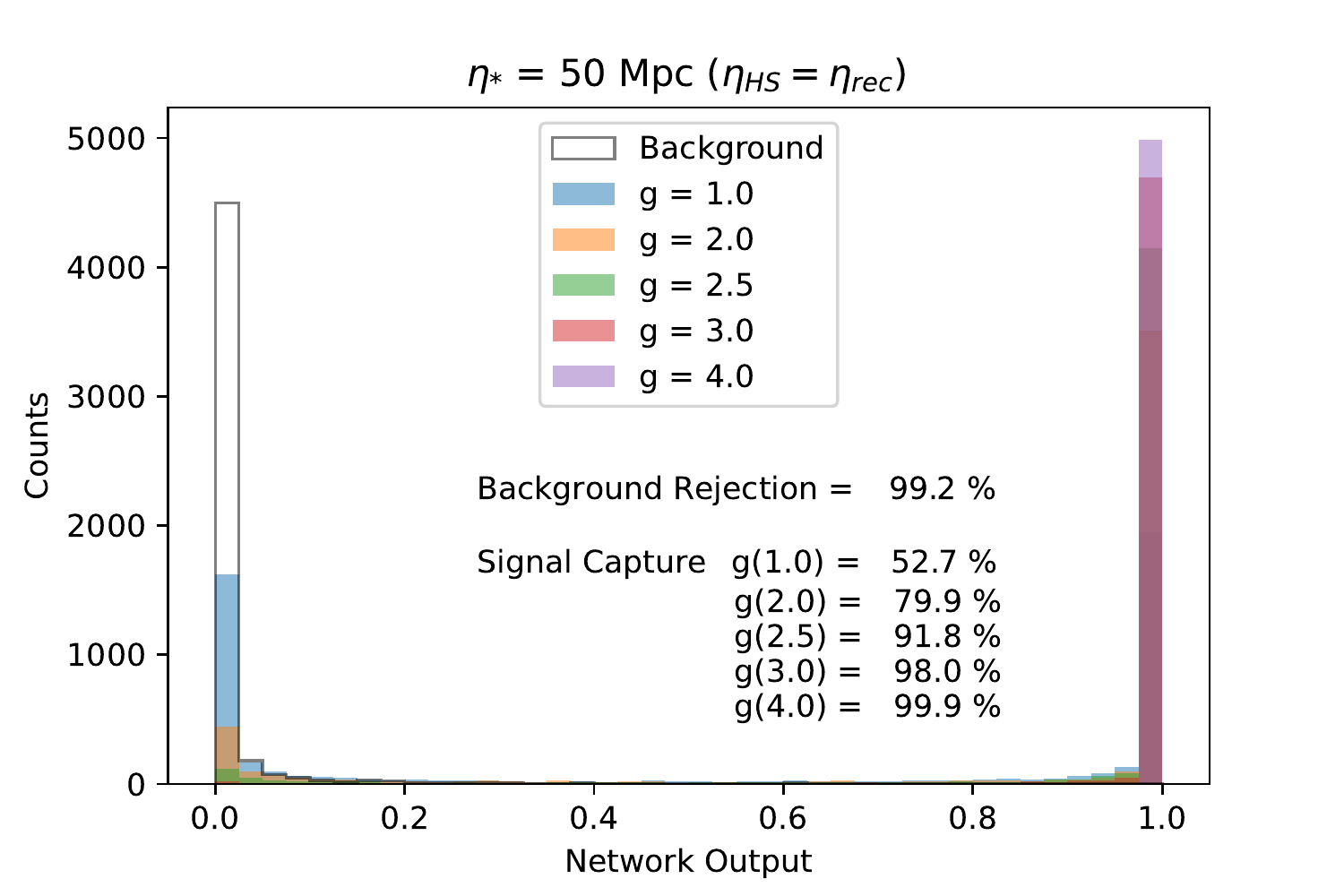} 
    \end{center}
    \vspace{-0.5cm}
    \caption{\label{fig:output_distrb} Network output for $5$k images without (blank histogram) and with (colored histograms) PHS signals. We count the image as an identified signal map when the network output $>0.5$. In the plots we show the background rejection rate from the CMB-only analysis and the signal capture rate from the CMB+PHS images, for different inflaton-$\chi$ couplings $g$. The fake rate is defined as $(1-$background rejection rate$)$. The plots on the left have both the hotspots distributed uniformly with separation $\leq\eta_*$ and within $\eta_{\rm HS}=\eta_{\rm rec}\pm\eta_*$, which is how we simulate the signal for the rest of the study. The signal capture rate therefore includes possible suppression due to hotspots moving off the last scattering surface. For comparison, we show the training results in the right plots requiring $\eta_{\rm HS}=\eta_{\rm rec}$. Comparing results obtained from the same study but with different sets of $5$k images, we find the efficiency numbers vary by $\sim 0.1 - 1\%$.}
\end{figure}

In Fig.~\ref{fig:output_distrb}, we show the network output for the $5$k images with and without injecting the PHS signal. 
In the left column we show the result when the PHS are uniformly distributed within a shell of $\eta_{\rm rec}\pm \eta_*$ around the surface of last scattering, while the right column shows the result when $\eta_{\rm HS} = \eta_{\rm rec}$. 
The signal capture and background rejection rates in Fig.~\ref{fig:output_distrb} refer to $\epsilon_{S,90^2}$ and $(1-\epsilon_{B,90^2})$.
Clearly, for $g \ge 3$, our CNN setup is highly efficient at separating CMB+PHS images from CMB images alone. 
For example, for $g=3$ (the same coupling as in the training sample) and $\eta_* = 160\, \text{Mpc}$, $\epsilon_{S,90^2}$ is over $73\%$ with $\epsilon_{B,90^2}$ less than $0.1\%$.  
For $\eta_*=160$~Mpc and $100$~Mpc, the signal capture rate falls if the hotspots are off the last scattering surface but in the $\eta_{\rm rec}\pm\eta_*$ window we consider. 
When applying the same trained network on dimmer PHS signals ($g < 3$), $\epsilon_{S,90^2}$ drops, but the background rejection rate remains close to unity.

Both $\epsilon_{S, 90^2}$ and $\epsilon_{B,90^2}$ vary with the horizon size. 
Comparing results for $\eta_* = 160\,\text{Mpc}$ to $\eta_* = 50\, \text{Mpc}$, the $\epsilon_{S,90^2}$ values are similar for $g \ge 3$, but $\eta_* = 50$~Mpc case performs much better at weaker coupling ($\epsilon_{S,90^2} = 51.2\%$ for $\eta_* = 50\,\text{Mpc}$ compared to $1.8\%$ for $\eta_* = 160\,\text{Mpc}$, both for $g = 1$). 
The $\eta_* = 50\, \text{Mpc}$ case has a larger background fake rate, compared to $\eta_* = 160$~Mpc.  
However, even if we incorporate the background and compare $\epsilon_{S, 90^2}/\sqrt{\epsilon_{B,90^2}}$ -- the efficiency ratio is $\mathcal O(10)$ times larger for the dimmer, $\eta_* = 50$~Mpc case. The ability of catching dimmer signals indicates that the network uses additional information than the overall temperature to identify the PHS.

Although it is difficult to know exactly how the CNN identifies the PHS, the network seems to more accurately identify PHS with a distinct rim structure compared to just utilizing the fact that there are two hotspots (Fig.~\ref{fig:PHSsample}). 
One indication that the CNN utilizes the rim structure of the $\eta_*=50$~Mpc signal is that the signal capture rate for that benchmark is insensitive to whether or not the PHS lie on the last scattering surface. 
We perform the same CNN analysis by having the signal hotspots centered on the last scattering surface ($\eta_{\rm HS}=\eta_{\rm rec}$ in Eq.~(\ref{eq.thprofile})) and summarize results in the right column of Fig.~\ref{fig:output_distrb}. 
For hotspots with temperature profile peaked at center, as we show in the $\eta_*=160$ and $100$~Mpc plots in Fig.~\ref{fig:delTvsrad}, the highest PHS temperature takes the maximum value when $\eta_{\rm HS}=\eta_{\rm rec}$ (orange). 
It then is reasonable to have a larger average signal capture rate when the hotspots center on the last scattering surface. 
However, as we illustrate in the upper left plot in Fig.~\ref{fig:delTvsrad}, the ``shell" of the $\eta_*=50$~Mpc signal in 3D always project into a rim with a fixed temperature (at angle $\approx 0.008$~rad), regardless of the location of the hotspot, $\eta_{\rm rec}, \eta_{\rm rec}+\eta_*$, or $\eta_{\rm rec}-\eta_*$. 
Therefore, if the CNN identifies the $\eta_*=50$~Mpc signal based on the rim structure, $\epsilon_{S, 90^2}$ should remain the same even when the PHS are on the last scattering surface. 
This is indeed what we see on the bottom plots in Fig.~\ref{fig:output_distrb}. 
Further study on what features the CNN uses to identify the $\eta_* = 50\, \text{Mpc}$ case can be found in Appendix~\ref{app.50shape}.

\subsection{Application of the Trained Network to Larger Sky Maps}

After training the CNN to identify PHS in images with $90^2$ pixels, we look for signals on a larger sky map by applying the same network analysis repeatedly across the larger map. 
In this way we can analyze, in principle, arbitrarily large maps.  
A benefit of such a larger map search is that it avoids the loss of sensitivity to signals where a PHS is partially cut out by the boundary of a $90^2$-pixels region. 
Such a PHS would be lost had we simply partitioned the sky into non-overlapping $90^2$-pixels regions. 

For a concrete application, we study maps with $720^2$ pixels\footnote{Repeating this analysis on even larger maps would be ideal and lessen the assumptions made when extrapolating to the whole sky. However, practically, we found that $720^2$ was the largest size we could make without sacrificing statistics (number of maps).} using the following steps: (i) we apply the trained network on the upper left corner of the map, obtaining the network output, (ii) we shift the $90^2$-pixels ``window'' to the right by $5$ pixels and get the network output again, (iii) repeat the process until we hit the right hand side of the large map. Then, return to the upper left corner but slide the widow down by $5$ pixels, (iv) continue with these steps until the entire larger map is covered. 
The result of steps (i) - (iv) result in what we call a ``probability map''. 
Starting with an original $720^2$ image and scanning in steps of $5$ pixels, the probability map has $126^2$ entries, with each entry showing the probability of having a signal in a $90^2$-pixels region centered at each pixel. 
We have tried different step sizes and find that a $5$ pixel step size yields nearly identical results to a $1$ pixel step size for the following analysis, so we use the $5$ pixel step size for improved computational speed. 

\begin{figure}[!t]
    \begin{center}
    \includegraphics[width=0.4\textwidth,clip]{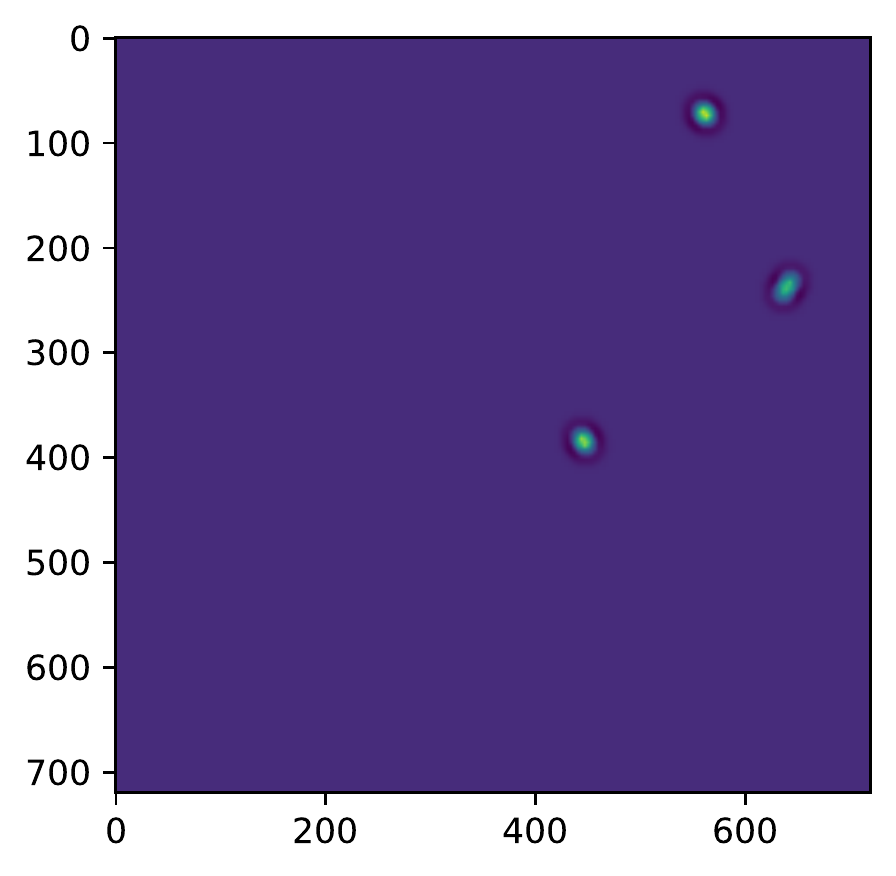}
   \quad
    \includegraphics[width=0.4\textwidth,clip]{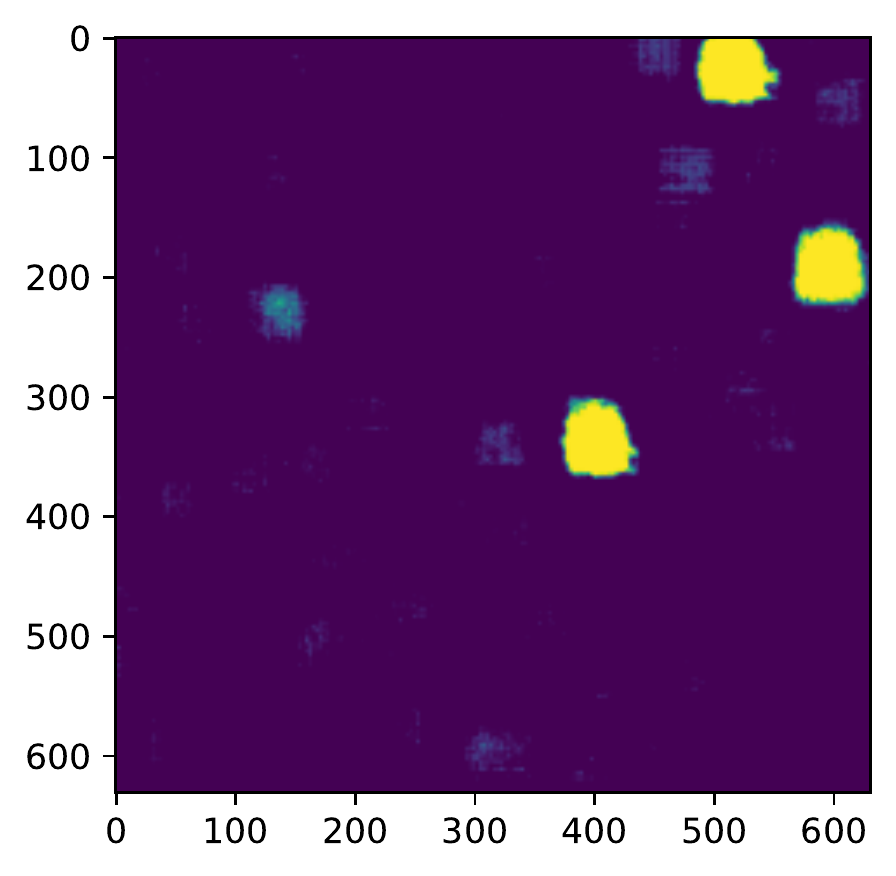}
    \end{center}
    \vspace{-0.1cm}
    \caption{\label{fig:10skip_scan} \emph{Left}: PHS signals that are implanted on CMB map. \emph{Right}: Probability map from scanning the same $720^2$ image plus the CMB with the CNN search of $90^2$-pixels region shifting in steps of 1 pixel. The true and fake signals show up as clusters in the processed image. We further suppress the number of fake signals in the following analysis by applying cuts on the network output of each pixel and the pixel number in each cluster. We find the analysis from shifting the search window in steps of 5 pixels produce similar results to the steps of 1 pixel and therefore use 5 pixel steps for the rest of the analysis.}
\end{figure}

In Fig.~\ref{fig:10skip_scan}, we show an example of the probability map (right) obtained from the image of CMB plus three PHS signals shown in the left plot. 
To make the signals more visible by eye, we make the probability map by sliding the search window in $1$ pixel step, thereby giving a $630^2$-pixels map. 
The true signals in the right plot show up as three bright clusters, and there are fake signals from the CMB fluctuations themselves. 
To reduce these fake signals, we further apply cuts on the probability map by only keeping pixels satisfying a threshold cut (network output) $>0.9$. 

To properly count the number of observed signals, we cluster nearby pixels in the probability map. 
Specifically, we employ the ``scikit-image's morphology label'' technique \cite{van2014scikit} to connect neighboring pixels with the same values and therefore group the connected pixels into clusters. 
We further remove clusters with $<30$ connected pixels, as these have a smaller chance to be a true signal. 
The choices of cuts on network output threshold and number of pixels in a cluster were made by trying several values and choosing the value that maximized the signal capture rate while keeping the fake rate small.

To determine how well the method works, we study a set of $500$ CMB-only images (i.e., pure background) and $500$ images where $N_{\rm inj}=3$ randomly distributed PHS signals have been injected. We then define `efficiencies': 
\begin{align}
\epsilon_{S,720^2} = \frac{\text{total number of clusters passing the threshold in CMB+PHS maps}}{3\times500} \nonumber \\
\epsilon_{B,720^2} = \frac{\text{total number of clusters passing the threshold in CMB-only maps}}{500}.
\end{align}
The results for the benchmark $\eta_*$ values and three different couplings are summarized in Table~\ref{tab.720offLSS} below. The $720^2$ search retains some of the key features from the $90^2$ search, namely the superior performance for $\eta_* = 50\, \text{Mpc}$ when $g$ is small.

\begin{table}[h!]
\begin{tabular}{|c|c|c|c|}
\hline
          & $\eta = 50$ Mpc & $\eta = 100$ Mpc & $\eta = 160$ Mpc \\ \hline
$\epsilon_{B,720^2}$ & 1.4 \%          & 11 \%            & 6.6 \%           \\ \hline
$\epsilon_{S,720^2},\, g = 1$   & 54.6 \%           & 0.8 \%           & 0.5 \%             \\ \hline
$\epsilon_{S,720^2},\, g = 2$   & 84.0 \%         & 34 \%          & 34.6 \%          \\ \hline
$\epsilon_{S,720^2},\, g = 3$   & 98.6 \%         & 76.8 \%          & 71.2 \%          \\ \hline
\end{tabular}
\caption{CNN result from scanning $500$ randomly generated CMB or CMB+PHS maps using the network trained in Sec.~\ref{sec.training}. The image size is $720^2$ pixels, and we shift the search window having $90^2$-pixels by 5 pixel steps. The fake rate is the average number of fake signals from a $720^2$-pixels map with CMB-only. The signal capture rate is the chance of identifying each input PHS signal. Comparing results obtained from the same study but with different sets of $500$ images, we find the efficiency numbers vary by $\sim 0.1 - 1\% $. }\label{tab.720offLSS}
\end{table}

\subsection{Obtaining Theoretical Bounds from Detection Statistics}

Using Table~\ref{tab.720offLSS} we can calculate the upper bound on $N_{\rm PHS}$, the number of hotspot pairs produced where both members of a pair lie within a shell of $\pm\eta_*$ from the last scattering surface. 
As an example, let us take $\eta_* = 50$~Mpc and $g=1$. 
From Table~\ref{tab.720offLSS}, we see $\epsilon_{S,720^2} = 54.6\%$ while $\epsilon_{B, 720^2}=1.4\%$. 
Assuming that only a fraction $f_{\rm sky} = 60\%$ is used for the search, the total number of signals for this benchmark is $Sig = \epsilon_{S,720^2}\, N_{\rm PHS}\, f_{\rm sky}$, while the number of background events is $Bg = 25\, \epsilon_{B,720^2}\, f_{\rm sky}$, where the factor of $25$ is the number of $720^2$ patches needed to cover the full sky. 
From the number of signal and background events, we form the log-likelihood ratio~\cite{Cowan:2010js,Craig:2015jba} and then solve for $N_{\rm PHS}$ for the desired signal significance. When calculating the $2\sigma$ exclusion bound, we require
\beq
  \sigma_{exc} \equiv
    \sqrt{-2\,\ln\bigg(\frac{L(Sig\!+\!Bg | Bg )}{L( Bg | Bg)}\bigg)} \geq 2 ,
  \;\;\;\;\; \text{with}\;\;\;
  L(x |n) =  \frac{x^{n}}{n !} e^{-x}\,.
\label{Eq:SigExc} 
\eeq
Note that this is the expected bound, as we are taking simulated CMB background to be the number of observed events ($n$ in Eq.~\eqref{Eq:SigExc}). 
The resulting values of $N_{\rm PHS}$ are given in the left panel of Table.~\ref{tab.result}. It is also interesting to determine how many PHS would be needed for discovery at each benchmark point. We calculate the expected discovery reach using
\beq
  \sigma_{dis} \equiv
    \sqrt{-2\,\ln\bigg(\frac{L(Bg | Sig\!+\!Bg)}{L( Sig\!+\!Bg| Sig\!+\!Bg)}\bigg)} \geq 5 \,.
\label{Eq:SigDis} \eeq
The results are collected in Table~\ref{tab.result2}.

\begin{figure}
    \centering
    \begin{subfigure}[b]{0.328\textwidth}
\includegraphics[width=\textwidth, clip]{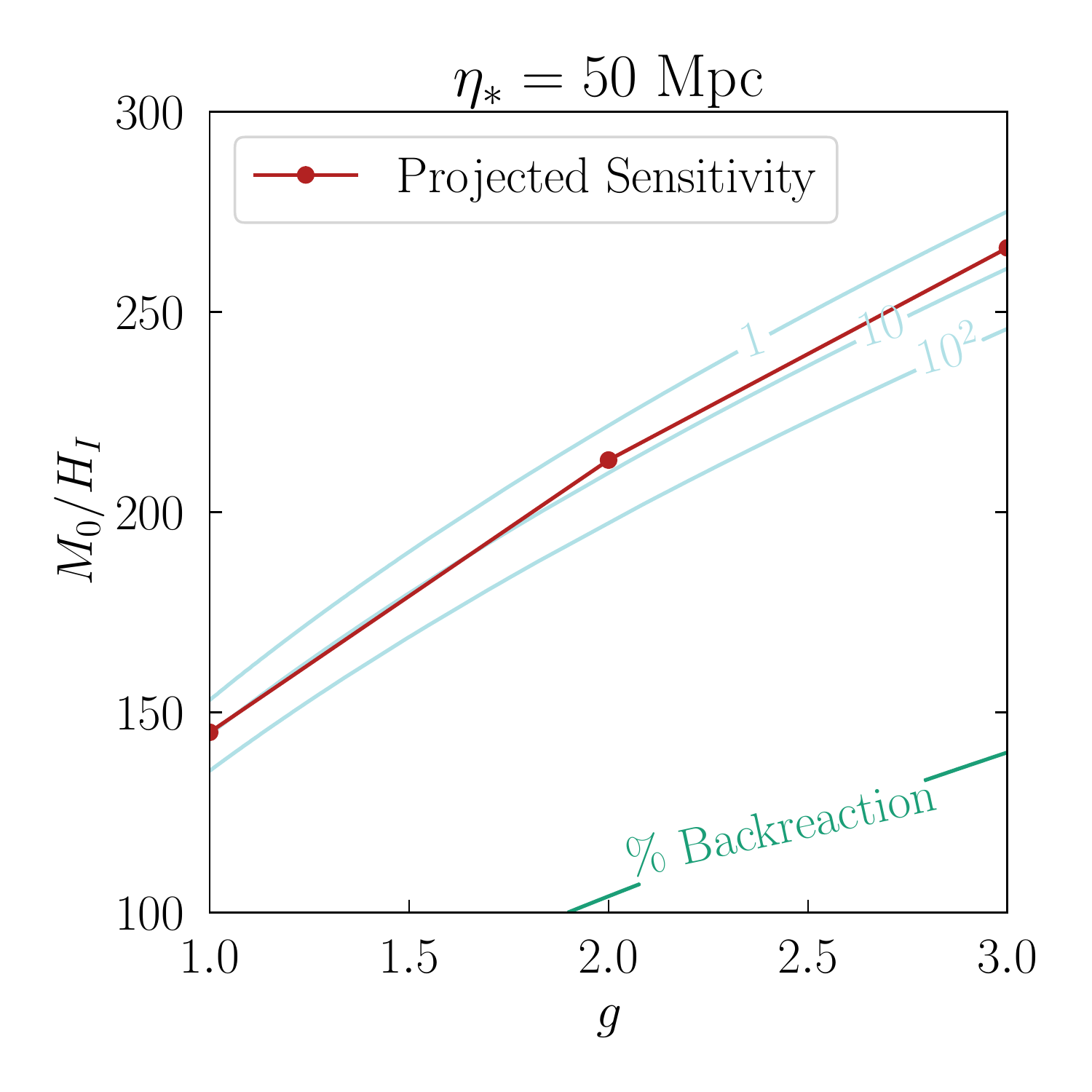}
    \end{subfigure}
    \vspace{0.1mm}
    \begin{subfigure}[b]{0.328\textwidth}\includegraphics[width=\textwidth,clip]{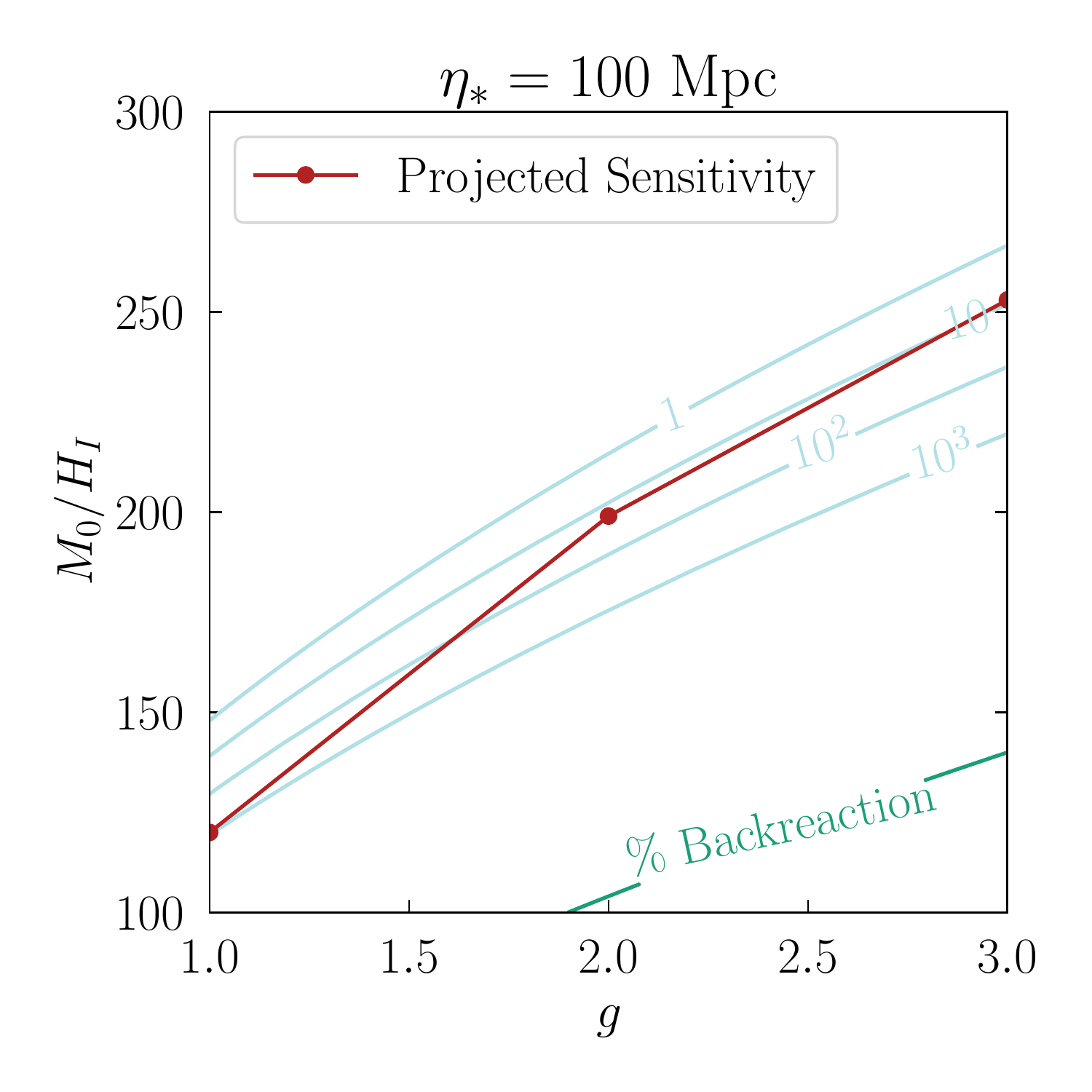}
    \end{subfigure}
    \vspace{0.1mm}
    \begin{subfigure}[b]{0.328\textwidth}\includegraphics[width=\textwidth,clip]{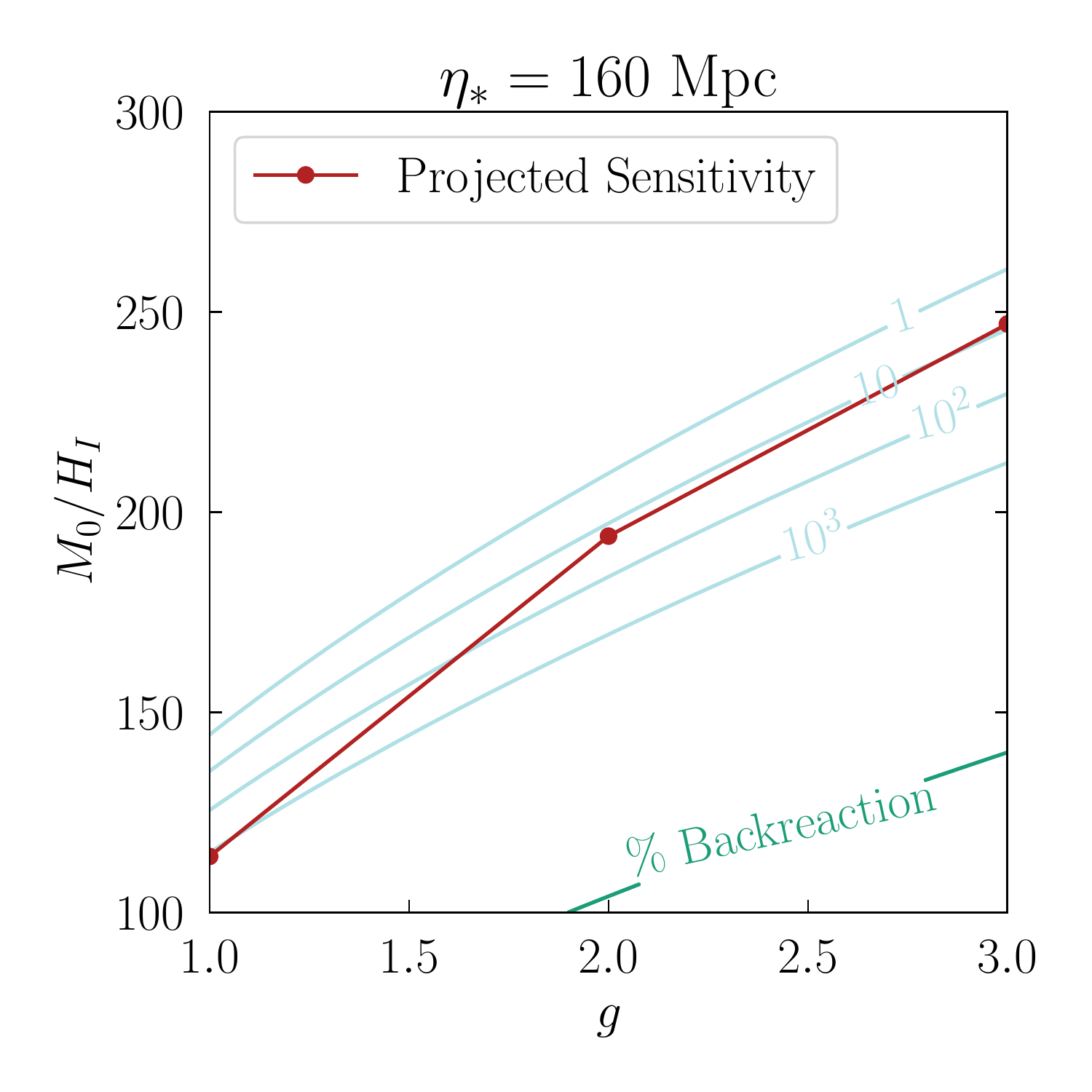}
    \end{subfigure}
    \caption{Bound on the heavy scalar mass for $\eta_*=50$~Mpc, $\eta_*=100$~Mpc, and $\eta_*=160$~Mpc. In the region above the `$\%$ Backreaction' line, the backreaction to the inflationary dynamics due to particle production is smaller than a percent (see Ref.~\cite{Kim:2021ida} for a more detailed discussion). The light blue lines show various contours of $N_{\rm PHS}$. We notice that the projected CNN search is able to cover most of the parameter space up to the target $N_{\rm PHS}=1$ contour.} 
    \label{fig:my_label}
\end{figure}

We can further obtain the minimum mass $M_0$ of the heavy particle corresponding to $\sigma_{exc}$ and $\sigma_{dis}$ using Eq.~(\ref{eq.Nspots}) and $\Delta\eta=2\eta_*$.\footnote{One subtlety in solving the mass bound is that when simulating the PHS signals, we require both hot spots to be within $\pm\eta_*$ around the last scattering surface. Hence, the simulation excludes PHS with one of the hot spots outside of the shell region that would be harder to see by the CNN. However, when solving the upper bound on the PHS density using Eq.~(\ref{eq.Nspots}), we take into account the signals that are partially outside of the shell region, leading to an over-estimate of the signal efficiency and a stronger upper bound on the number density. From checking the hot spot distribution numerically, we find that $\approx 17\%$ of the PHS in our examples can be partially outside of the $\pm\eta_*$ region. Fortunately, since the size of $M_0$ only depends on the number density bound logarithmically, the error only changes the $M_0$ bound by up to $1\%$. This is acceptable for the accuracy we want for the concept study.} 
In Table~\ref{tab.result} and~\ref{tab.result2}, we show the bounds (or reach) on the number of PHS and $M_0/H_I$. 
Due to the energy injection from the dynamics of the inflaton, we can probe scalar particles with masses up to $\approx 260 H_I$.  
In the bottom right tables, we show that the mass bounds correspond up to $\approx2.6$ times the mass-changing rate caused by the inflaton rolling ($\sqrt{g\dot{\phi}_0}$ ), which dominates the exponential suppression in Eq.~(\ref{eq.Nspots}). We also plot the $2\sigma$ lower bound on $M_0/H_I$ in Fig.~\ref{fig:my_label}. Since the $N_{\rm PHS}$ depends on $M_0$ exponentially, a slightly lower scalar mass than the $2\sigma$ bound leads to a $5\sigma$ discovery of the PHS. 

These bounds are significantly improved compared to the previous analysis in Ref.~\cite{Kim:2021ida}; this is not surprising given that the analysis in Ref.~\cite{Kim:2021ida} was very simplistic, utilizing only a single temperature cut to separate signal from background. 
Using the CNN, we can now obtain meaningful bounds for $g=1,\,2$ -- cases for which the PHS were rather invisible before. For hotter signals, e.g. $g=3$, the CNN analysis beats the past result by $\Delta M_0\approx 60H_I$. This is a notable improvement given that the PHS density is exponentially sensitive to the scalar mass (squared).

Finally, to show that the CNN search of localized objects gives a better probe of heavy particle production than the measurement of CMB temperature power spectra, we plot the corrections to the $\Lambda$CDM $D_{\ell}^{\rm TT}$ spectrum in Appendix~\ref{app.DlTT}, including the same number of PHS in Table.~\ref{tab.result}. 
For example, for $g=1$,\, $\eta_*=160$~Mpc, we see from Table~\ref{tab.result} that the $2\sigma$ bound on $N_{\rm PHS}$ from our CNN analysis is 1162 hotspot pairs. 
Injecting 1162 hotspots into the sky,\footnote{For simplicity, we restrict all hotspots to the last scattering surface. This somewhat overemphasizes the PHS correction to the power spectrum, as scenarios with both particles fixed to the last scattering surface are, on average, brighter than when $\eta_{HS}$ varies.} we find a correction to $\mathcal D_{\ell}^{\rm TT}$  of $\Delta\chi^2 = 0.3$ -- well within the $1\sigma$ band on Planck 2018 temperature power spectrum. 
Repeating this exercise with the other benchmarks in Table \ref{tab.result}, yields $\Delta\chi^2$ values that are even smaller.

\begin{table}[!t]
\begin{tabular}{|c|c|c|c|}
\hline
 Number of PHS       & $\eta = 50$ & $\eta = 100$ & $\eta = 160$ \\ \hline
$g = 1$ & 8     &840     & 1162                \\ \hline
$g = 2$ & 5    &20      & 17                \\ \hline
$g = 3$ & 4     &9     & 8                    \\ \hline
\end{tabular}
\\\vspace{1em}
\begin{tabular}{|c|c|c|c|}
\hline
 $M_0/H_I$       & $\eta = 50$ & $\eta = 100$ & $\eta = 160$ \\ \hline
$g = 1$ & 145     &120    & 114                   \\ \hline
$g = 2$ & 213    &199      & 194                \\ \hline
$g = 3$ & 266    &253     & 247                   \\ \hline
\end{tabular}
\quad
\begin{tabular}{|c|c|c|c|}
\hline
 $M_0/(g\dot\phi_0)^{1/2}$       & $\eta = 50$ & $\eta = 100$ & $\eta = 160$ \\ \hline
$g = 1$ & 2.5    &2.0    & 2.0                   \\ \hline
$g = 2$ & 2.6    &2.4    & 2.4                \\ \hline
$g = 3$ & 2.6    &2.5    & 2.4                   \\ \hline
\end{tabular}
\caption{\emph{Upper:} $2\sigma$ upper bound on the number of PHS in the whole CMB sky with both hotspot centers located within $\eta_{\rm rec}\pm\eta_*$ window around the last scattering surface. In the calculation we assume sky fraction $f_{\rm sky}=60\%$. \emph{Lower left}: lower bounds on the bare mass of the heavy scalar field in units of the Hubble scale during the inflation. \emph{Lower right}: lower bounds on the bare mass in units of the rate of the mass, $(g\dot\phi_0)^{1/2}$, owing to the inflaton coupling.}\label{tab.result}
\end{table}

\begin{table}[!t]
\begin{tabular}{|c|c|c|c|}
\hline
 Number of PHS       & $\eta = 50$ & $\eta = 100$ & $\eta = 160$ \\ \hline
$g = 1$ & 16     &2047     & 2757                \\ \hline
$g = 2$ & 10    &48      & 40                \\ \hline
$g = 3$ & 9     &21     & 19                    \\ \hline
\end{tabular}
\\\vspace{1em}
\begin{tabular}{|c|c|c|c|}
\hline
 $M_0/H_I$       & $\eta = 50$ & $\eta = 100$ & $\eta = 160$ \\ \hline
$g = 1$ & 143     &116    & 110                   \\ \hline
$g = 2$ & 210    &194      & 189              \\ \hline
$g = 3$ & 262    &247     & 241                   \\ \hline
\end{tabular}
\quad
\begin{tabular}{|c|c|c|c|}
\hline
 $M_0/(g\dot\phi)^{1/2}$       & $\eta = 50$ & $\eta = 100$ & $\eta = 160$ \\ \hline
$g = 1$ & 2.4    &2.0    & 1.9                  \\ \hline
$g = 2$ & 2.5    &2.3    & 2.3                \\ \hline
$g = 3$ & 2.6    &2.4    & 2.4                   \\ \hline
\end{tabular}
\caption{Same as Table~\ref{tab.result} but for the $5\sigma$ discovery reach. }\label{tab.result2}
\end{table}

\subsection{Comparison with a Matched Filter Analysis}\label{sec.MF}
Matched filter analysis is a standard tool for identifying localized signals on a CMB map. 
Given a 2D power spectrum of the CMB, $P(k)$, we can obtain a filtered map $\psi(\vec{r})$ in position space from a convolution between the original image (signal plus background) $\zeta(\vec k)$ and a signal filter $h(\vec k)$ (the Fourier transform of a profile $h(\vec{r})$ in position space),
\beq\label{eq.MF}
\psi (\vec{r}) = \int\frac{d^2\vec{k}}{(2\pi)^2}\left( \frac{\zeta(\vec k) h(\vec k)}{P(k)} \right)\,e^{i\vec{k}\cdot\vec{r}}.
\eeq
If the signal is spherically symmetric, the filter simplifies to $h(\vec k)=h(k)$. From the filtered map $\psi (\vec{r})$ one can construct an optimal likelihood ratio test between the Gaussian null hypothesis and the existence of the signal (see e.g. \cite{Munchmeyer:2019wlh}), making the matched filter ideal for picking out single (or more generally, non-overlapping) localized signals. 

As we have seen, while the individual hotspots are spherically symmetric, they often overlap (at least for the range of parameters we are interested in), leading to a net signal in the sky that is no longer spherical. Additionally, the random separation between the initial heavy particles means the resulting PHS are not uniform. The unusual shape and variability among signals make the PHS less suitable for a vanilla matched filter analysis. While it may be possible to design a complicated and large bank of matched filters to cover the space of possible signal templates, the CNN analysis can effectively learn a set of flexible filters to enhance the signal over background even with varying and non-spherical signal shapes.


\begin{figure}[t!]
    \begin{center}
    \includegraphics[width=0.32\textwidth,clip]{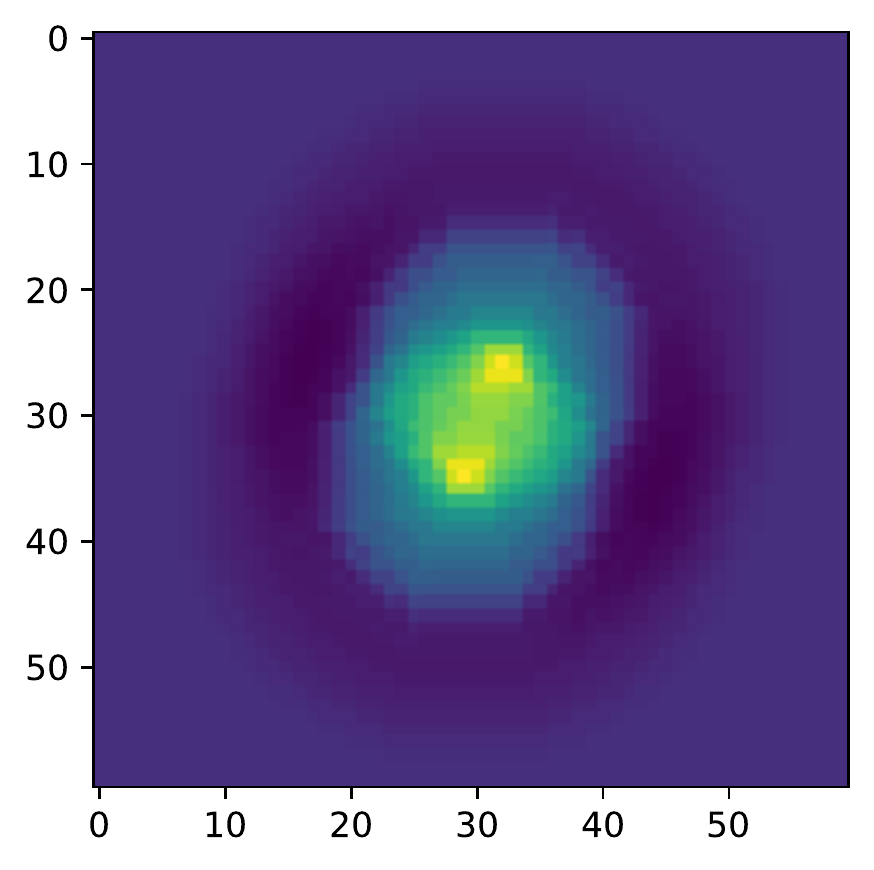}\,
    \includegraphics[width=0.32\textwidth,clip]{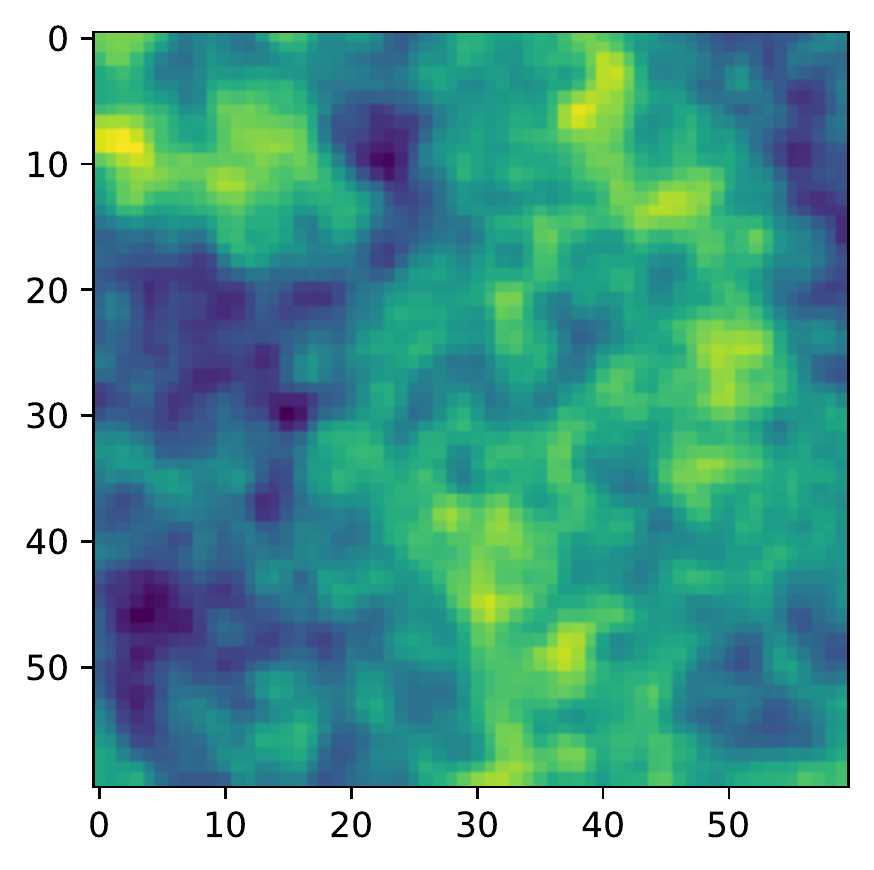}\,
    \includegraphics[width=0.32\textwidth,clip]{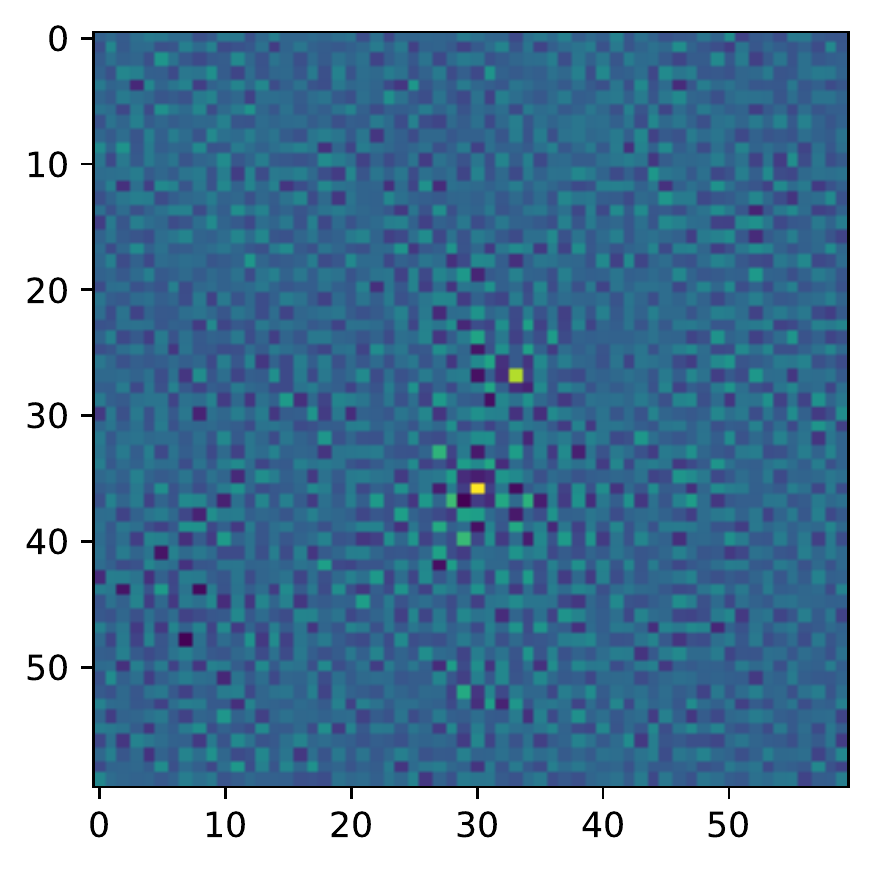}
    \end{center}
    \vspace{-0.5cm}
\caption{\label{fig:MFpic} Example images from the matched filter analysis. \emph{Left:} PHS with $\eta_*=160$~Mpc and $g=2$. \emph{Middle:} Signal plus the background. \emph{Right:} Filtered map from the convolution integral Eq.~(\ref{eq.MF}).}
\end{figure}

Even if the matched filter analysis defined in Eq.~(\ref{eq.MF}) is not optimal for the full pairwise hotspot signal, it is still instructive to compare a few examples of the matched filter analysis versus the CNN. For this comparison, we consider PHS that lie only on the last scattering surface. The combined signal from the PHS will still be non-spherical, but restricting all PHS to the last scattering surface does take away some of the variability among signals.\footnote{We still allow a random separations (within $\eta_*$) between hotspots {\em on} the (2D) last scattering surface} While each hotspot in a pair will ``pollute'' the other -- meaning that it appears as a background that is different from the CMB fluctuations -- each of the two hotspots can still be picked up effectively by the single spot template $h(k)$.

We perform the comparison using $90^2$ pixel images with one PHS injection. We use \Quicklens to generate the CMB maps, which follows periodic boundary condition and thereby ensures the separation between $k$-modes in the 2D power spectrum $P(k)$ of the CMB image. The CNN results for this signal set have already been shown in Sec.~\ref{sec.training} and can be found in the right hand panels of Fig.~\ref{fig:output_distrb}; the background rejection is above $99\%$ for all benchmark points, while the signal capture rate varies from a few percent to $100\%$ depending on $\eta_*$ and $g$.

For the matched filter analysis, we obtain $P(k)$ from the average of the discrete Fourier transform of $500$ simulated images. We also apply discrete Fourier transform on the profile of a single hotspot in the PHS, and use it as $h(k)$ in the convolution. Carrying out the integral in Eq.~(\ref{eq.MF}), we obtain the processed maps $\psi(\vec{r})$. An example of the signal processing is shown in Fig.~\ref{fig:MFpic}, where the plot on the left is the PHS signal ($\eta_*=160$~Mpc and $g=2$), the middle is the signal plus background, and the right plot is the output image $\psi(\vec{r})$. We see that the filter can indeed pick up the signal hidden inside the background. 

\begin{figure}[!t]
    \begin{center}
    \includegraphics[width=0.49\textwidth,clip]{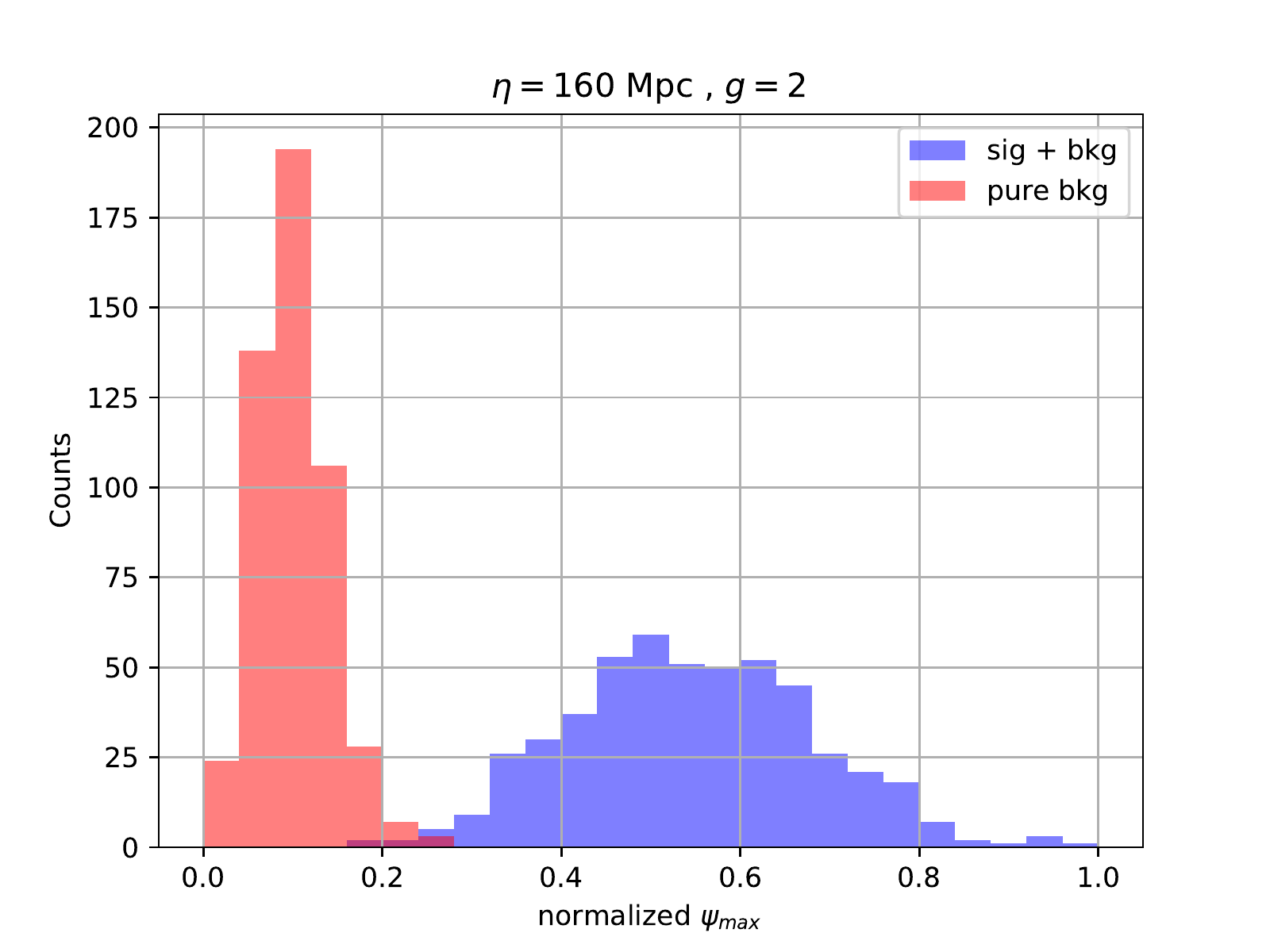}
    \includegraphics[width=0.49\textwidth,clip]{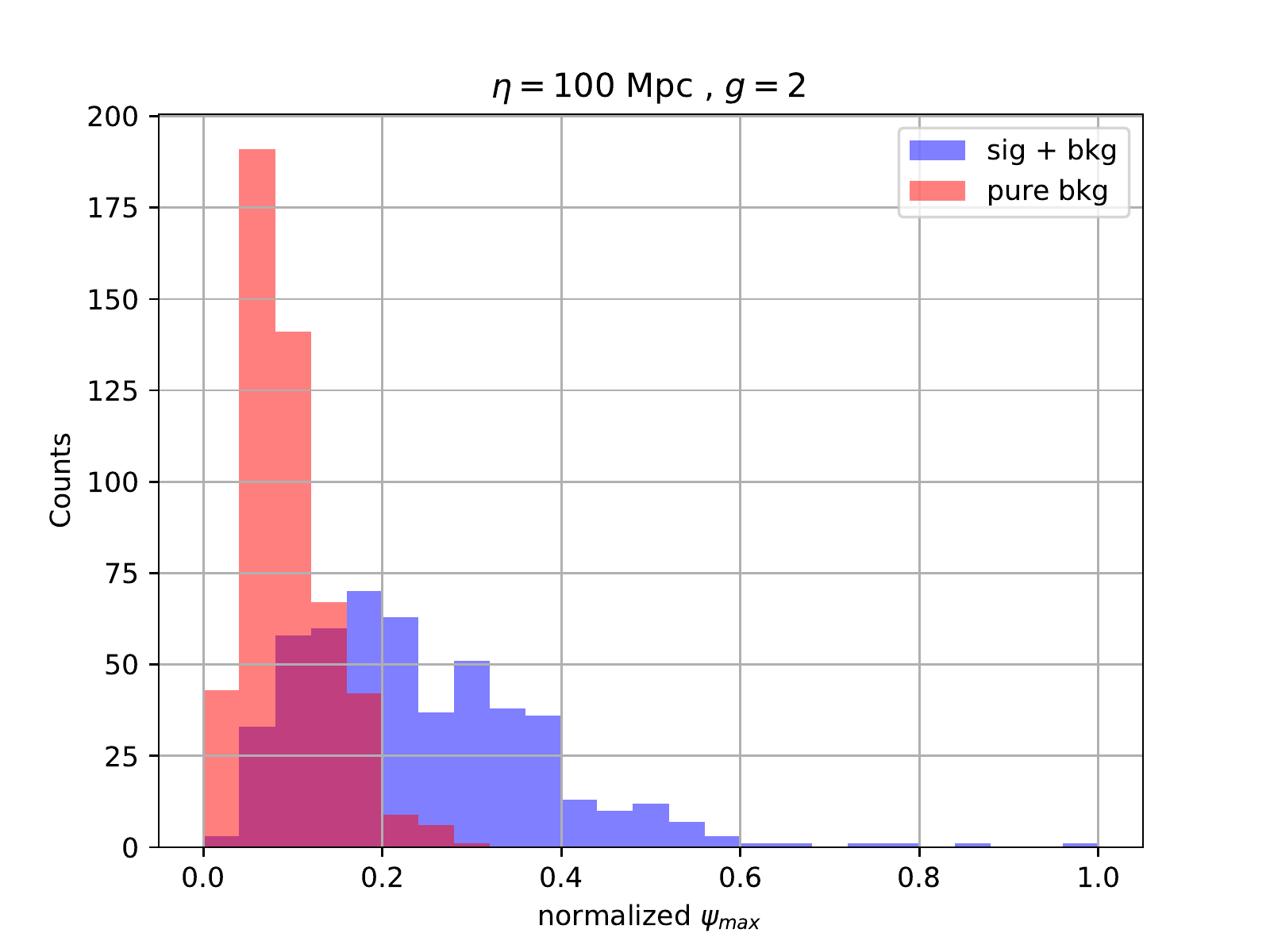}
    \end{center}
    \vspace{-0.2cm}
    \caption{\label{fig:MFdistg2} Maximum pixel distribution in filtered maps, where the value $\psi$ of the pixels on the filtered map is defined in Eq.~(\ref{eq.MF}). We use 500 CMB-only and 500 CMB+PHS maps and plot the distribution of the maximum $\psi$ of each filtered map to show the separation between the CMB and CMB+PHS results. We used feature scaling also known as min-max normalization for $\psi_{max}$, so that the smallest value is zero and the largest value is 1.}
\end{figure}

As one way to quantify the matched filter results, in Fig.~\ref{fig:MFdistg2} we show the distribution of largest $\psi(\vec{r})$ values in each of the $500$ maps generated with (blue) and without (red) PHS signals with $\{\eta_*,g\}=\{160\,{\rm Mpc},2\}$ (left) and $\{100\,{\rm Mpc},2\}$ (right). From this perspective, the matched filter clearly separates the signal and background for the two cases. We also perform the same analysis for the $\eta_*=50$~Mpc signals (which have much lower temperatures). In this case, the overlap between signal and background in the $\psi$ distribution is large, and a simple $\psi$ cut is not the optimal way to separate the signal and background. For this reason, we only consider the $\eta_*=100$ and $160$~Mpc examples in the following discussion. 

To provide a rough numerical comparison between the matched filter and the CNN analysis, we apply a  $\psi_{\rm max}$ cut in each of the matched filter histograms in Fig.~\ref{fig:MFdistg2}. We choose the $\psi_{\rm max}$ cut value to equal the background rejection rate in the CNN analysis, then compare signal capture rates in the two analyses. For the $\eta_*=160$~Mpc example, the signal capture rate is about $5\%$ and $74\%$ for $g=1$ and $2$, while the match filter analysis performs slightly better, capture rates $8\%$ and $98\%$ respectively.  For $\eta_*=100~\text{Mpc}$, the CNN signal capture rates are $\sim 10\%$ and $\sim 69\%$ for $g=1$ and $2$, while the match filter analysis rates are slightly lower, $4\%$ and $50\%$. 

In summary, we find that the CNN performs very close to the matched filter analysis, suggesting that it is near optimal. The advantage of the CNN, as we have discussed, is that it can learn to interpolate between all signal shapes that appear in our model.\footnote{We believe the small differences between the CNN and matched filter signal rates are due to the simplicity of the analysis -- where $\psi_{\rm max}$ is used as a proxy for the matched filter performance.}

\section{Discussion and Conclusion}\label{sec.concl}
In this work, we show that Convolutional Neural Networks (CNN) provide a powerful tool to identify pairwise hotspots (PHS) on the CMB sky.
These PHS can originate from superheavy particle production during inflation.
We improve the previous analysis of Ref.~\cite{Kim:2021ida} by more accurately modeling the distribution of PHS on the CMB sky and by developing a CNN-based signal search strategy. 
 
To accurately model the PHS distribution, we include the possibility that PHS are distributed along the line-of-sight direction, rather than fixed to the last scattering surface. As a result, the average inter-spot separation within a PHS, when projected onto the CMB, is smaller than in Ref.~\cite{Kim:2021ida}. 
For PHS with small values of $\eta_*$, such as $\eta_*= 50$~Mpc, the two hotspots in a PHS significantly overlap with each other, and the resulting PHS look like a single object, but with a distinct angular profile (Fig.~\ref{fig:PHSsample}).

For the signal search, we construct a CNN to identify PHS from within the CMB, the standard fluctuations of which act as backgrounds for the signal. 
The network is trained on $90^2$ pixel images with and without PHS injected in them (both with hotspots distributed in 3D, and with hotspots fixed on the last scattering surface). 
During training we choose a coupling $g=3$, but the trained CNN can still identify PHS for smaller values of $g$ with a significant signal capture rate and small background fake rate. 
We find that the CNN actually performs better for the smaller $\eta_*$ benchmark, even though the hotspots are dimmer. 
We believe this is due to the distinctive ring structure the PHS have when $\eta_* = 50\, \text{Mpc}$, as evidenced by comparing PHS signals distributed in 2D  versus in 3D, and by studies testing the CNN on `dot' and `ring' test signals (Appendix ~\ref{app.50shape}).

After developing the CNN for $90^2$ pixel images, we apply it to larger $720^2$ pixel maps, sliding $90^2$ `templates' in 5 pixel steps across the larger images to generate a probability map. 
In the probability map, each pixel is evaluated by the network multiple times. 
As a final step, we filter the probability map, only retaining clusters -- groups of positive network outcomes -- of a certain size. 
The benefit of the sliding template search is that it less sensitive to the exact position of the hotspot within the $90^2$ pixel region. Applied in this manner, we find that the CNN can efficiently discern the presence of hotspots, even if the signal temperature is much smaller than the CMB temperature fluctuations. In particular, the CNN can even identify $\mathcal{O}(10)$ number of PHS on the CMB sky for $g=1$ and $\eta_*=50$~Mpc, a signal that has a temperature $\approx20$ times colder than the average CMB temperature fluctuations. Translated into model parameters, for the benchmark models we study using mock CMB maps, we project that a CNN search can set a lower bound on the mass of heavy scalars $M_0/H_I\gtrsim 110-260$, with the precise value depending on the time of particle production and coupling to the inflaton. These numbers are a significant improvement over the simplistic analysis in Ref.~\cite{Kim:2021ida} that used single temperature cut to separate signal from the background.

Compared to the standard matched filter analysis, the CNN is more versatile in identifying non-rotationally symmetric signals with varying shapes and temperatures that arise in the context of PHS. We performed a simplified comparison between the CNN and matched filter analysis by considering PHS with a fixed profile and located on the last scattering surface to show that the match filter analysis can provide comparable signal capture and fake rates to the CNN search for PHS with $\eta_*=160$~Mpc and $100$~Mpc. For dimmer PHS ($\eta_*=50$~Mpc), more analysis is required to separate the signal and background in the filtered map. We leave a more detailed comparison to the matched filter method with a bank of filters to cover the signal space to future work.

Several future directions remain to be explored. It would be interesting to apply our methodology to actual Planck CMB maps to search for PHS. In the absence of a detection, we can still set a lower bound on the masses of ultra-heavy particles which are otherwise very difficult to discover or constrain. This, however, requires a subtraction of the astrophysical foregrounds and knowing if the CNN can distinguish PHS from the compact objects in the foreground. Since the distortion of the curvature perturbation from particle production also modifies structure formation at late times, it would also be interesting to see if the current or future Large Scale Structure (LSS) surveys can identify the resulting signals localized in position space. A neural network like the one used here can learn to incorporate the non-linear physics of structure formation if trained on suitable simulations. Related to localized PHS signatures, similar types of cosmological signals from topological defects~\cite{Cho:2014nka} or bubble collisions~\cite{Osborne:2013jea,Osborne:2013hea,McEwen:2012uk} can also arise and these may also be identified by a CNN search. From a more theoretical perspective, it would also be useful to write down a complete inflationary model that incorporates inflaton coupling to heavy fields and leads to particle production as described here. We leave these directions for future work.

\acknowledgments
We thank Raphael Flauger, Daniel Green, Matthew Johnson, Kin-Wang Ng, Bryan Ostdiek, LianTao Wang, Yiming Zhong for useful conversations. TK, AM, and YT are supported by the U.S. National Science Foundation (NSF) grant PHY-2112540. 
JK is supported in part by the National Research Foundation of Korea (NRF) grant funded by the Korea government (MSIT) (No. 2021R1C1C1005076), and in part by the international cooperation program managed by the National Research Foundation of Korea (No. 2022K2A9A2A15000153, FY2022).
SK is supported in part by the NSF grant PHY-1915314 and the U.S. Department of Energy (DOE) contract DE-AC02-05CH11231.
MM is supported by the U.S. Department of Energy, Office of Science, under Award Number DE-SC0022342.

\appendix

\section{Sensitivity to the $\Lambda$CDM Parameters}\label{app.LCDM}

Our analysis uses $\Lambda$CDM parameters in Eq.~(\ref{eq.default}) to simulate the CMB. As the $\Lambda$CDM parameters come with uncertainties, we should check how sensitive the signal capture rate is to the variation of the parameters. In Table~\ref{tab:diff_LCDMbkg}, we show the background rejection and signal capture rate using the same trained network for Fig~\ref{fig:output_distrb}~\emph{left} with $g=3$ and $\eta_*=160$~Mpc but on CMB maps simulated with variations of $\Lambda$CDM parameters. As we see, when changing the $\{A_s,\Omega_b,\Omega_{\rm CMB},n_s\}$ one by one with twice the $1\sigma$ uncertainty reported in~\cite{Aghanim:2018eyx}, the signal capture rate only changes by $\mathcal O(\text{few}\, \%)$, comparable to the variations in our CNN analysis due to finite sampling. The consistent search results show the robustness of the network's ability to identify PHS against the uncertainty of $\Lambda$CDM parameters.

\begin{table}[h!]
    \begin{tabular}{|c|c|c|c|c|c|c|c|c|}   
    
\hline
            & $\omega_b$ & $\omega_{\rm cdm}$   & $ 10^9A_s$ & $n_s$ & $\tau_{re}$  & Bg rejection & Sig capture \\ \hline\hline
Planck18  & 0.0224 & 0.120      & $2.10$ & 0.966  & 0.0543 &   $99.8\%$ & $74.0\%$      \\ \hline
Case 1 & & $+0.004$           &  &   &   & $99.8\%$ & $72.3\%$        \\ \hline
Case 2 & & & $+0.07$ &    &  & $99.2\%$ & $74.1\%$          \\ \hline
Case 3 &  & &   & $+0.01$  &  & $99.6\%$ & $69.9\%$           \\ \hline
Case 4 & $+0.0003$  & &   &   &  & $99.8\%$ & $73.4\%$           \\ \hline
Case 5 &  & &   &   & $+0.014$ & $99.2\%$ & $74.4\%$           \\ \hline
Case 6 & $+0.0003$ & $-0.004$     & $+0.05$ & $-0.01$  & $-0.014$ & $99.8\%$ & $72.4\%$        \\ \hline
\end{tabular}
    \caption{\label{tab:diff_LCDMbkg} The response of the signal capture and background rejection rates with varying $\Lambda$CDM parameters, labeled with the difference to the $\Lambda$CDM parameters. The variation of the rates is comparable to the fluctuations in our CNN analysis due to finite sampling and therefore is insignificant. For this test, we used $g =2$ and $\eta_*=160$~Mpc for the PHS signal.}
\end{table}

\section{PHS Corrections to the CMB Power Spectrum}\label{app.DlTT}
\begin{figure}[t!]
    \centering
    \includegraphics[width=0.45\textwidth,clip]{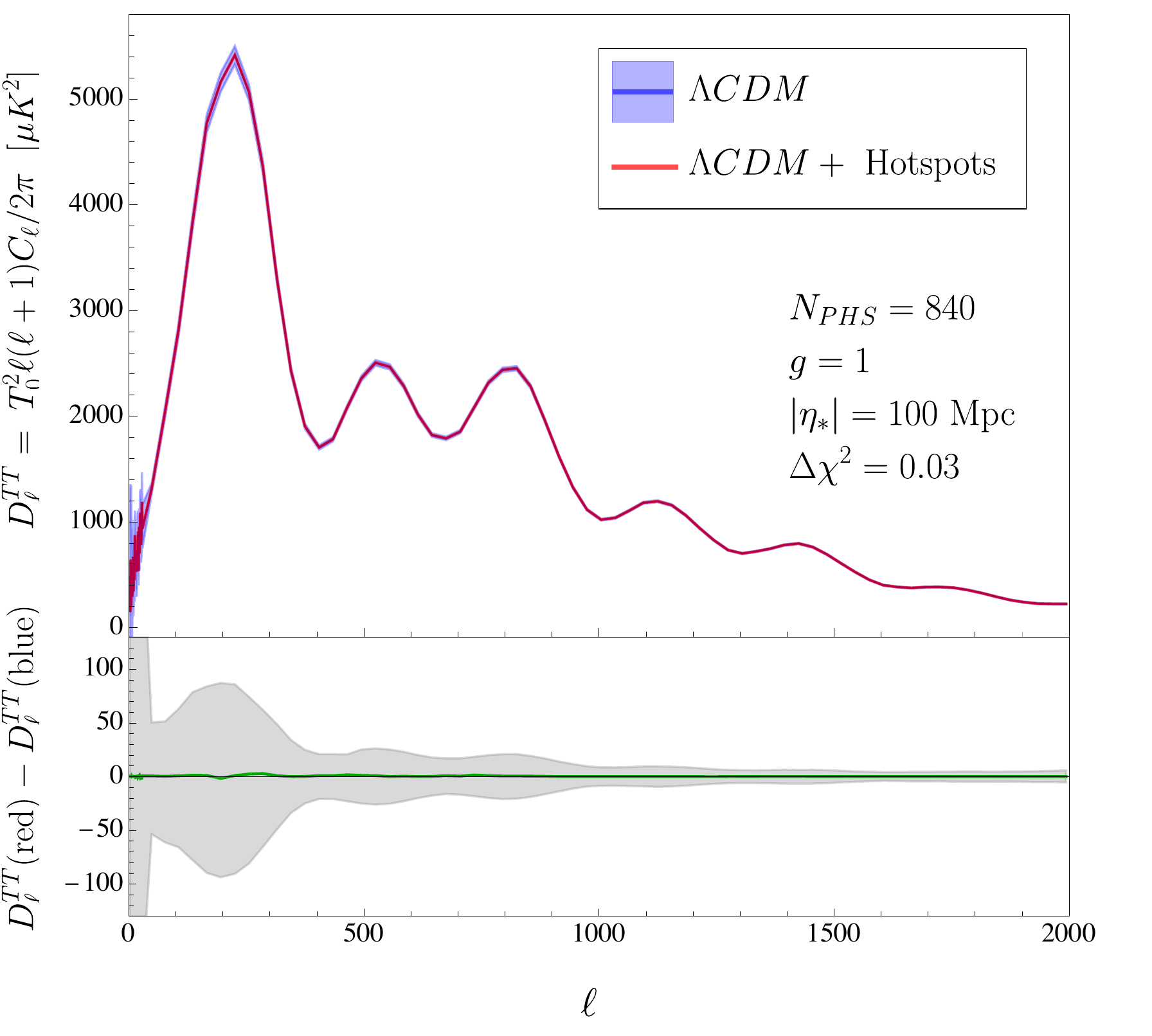}
    \includegraphics[width=0.45\textwidth,clip]{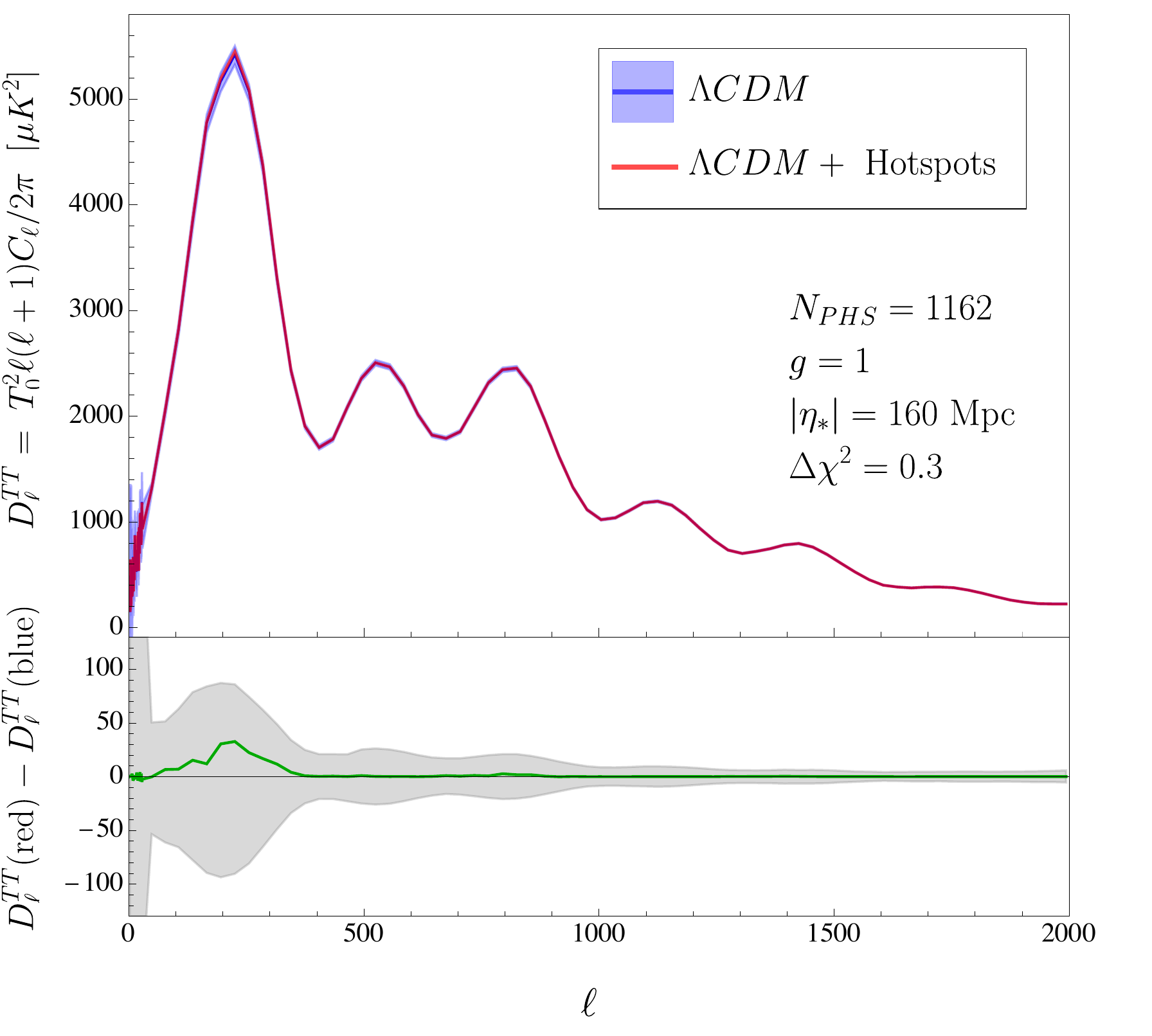}
    \caption{CMB temperature power spectrum using best fit $\Lambda \rm CDM$ input parameters in Eq.~(\ref{eq:params}) with (red lines) and without (blue lines) PHS signals implemented on the full sky using a resolution parameter $N_{\rm side} = 2048$.
    Here, we assume that all PHS signals are on the last scattering surface.
    The differences between the two distributions are shown in green lines, and the gray shaded regions denote $1\sigma$ uncertainty, taken from the Planck 2018 data.
    }
    \label{fig:PS}
\end{figure}

Here we show the corrections on the CMB power spectrum when the number of PHS in the full sky saturates the bounds in Table~\ref{tab.result}. We show examples with the coupling $g=1$ and horizon sizes $\eta_* = 100$~Mpc ($N_{\rm PHS}=840$) and $160~\text{Mpc}$  ($N_{\rm PHS}=1162$), assuming the centers of all the hotspots are located on the last scattering surface. Notice that the latter assumption of fixing $\eta_{\rm HS}=\eta_{\rm rec}$ makes the average PHS temperature higher compared to the main analysis that allows $\eta_{\rm HS}$ to vary. However, the assumption simplifies the power spectrum calculation and gives a more conservative result by exaggerating the PHS correction to the power spectrum. We also check results for different $g$ and $\eta_*$, but, following Table~\ref{tab.result}, with much smaller $N_{\rm PHS}$. The corrections to the power spectrum for the other benchmarks are even smaller.

To see how the excesses appear on the power spectrum, we utilize Hierarchical Equal Area isoLatitude Pixelization, \Healpix~\cite{Gorski:2004by}, based on the $C^{\rm TT}_{\ell}$ spectrum computed from the \CLASS~package using the same $\Lambda$CDM parameters in Eq.(\ref{eq:params}).
\Healpix~pixelates a sphere in an equal area where the lowest resolution consists of 12 baseline pixels. The resolution is increased by dividing each pixel into four partitions which can be parameterized as $N_{\rm pixels} = 12 N^2_{\rm side}$ where $N_{\rm side}$ is a power of 2. 
We choose the resolution parameter $N_{\rm side}=2048$.
Since the total number of pixels in a sphere characterizes the total number of independent $\ell$ modes in $C^{\rm TT}_{\ell}$, which is given by $\sum_{\ell=0}^{\ell_{\rm max}}(2\ell+1)=(\ell_{\rm max}+1)^2$, our benchmark resolution parameter $N_{\rm side}=2048$ corresponds to the maximum multipole number $\ell_{\rm max} \simeq 3500$.

Figure \ref{fig:PS} shows $\mathcal D_{\ell}^{\rm TT}$ spectra for the $\Lambda$CDM model (blue) and the $\Lambda$CDM+PHS (red) with $\eta_* = 100~\text{Mpc}$ and $\eta_* = 160~\text{Mpc}$.
The difference between the red and blue spectra is shown on the lower panel (green), with the $1\sigma$ error bar (gray) taken from the Planck~2018 result~\cite{Aghanim:2018eyx}. For both scenarios, the excesses are well below the error bar indicating that the power spectrum analysis will not be able to resolve them. We also show $\Delta\chi^2$ to quantify the deviations with respect to the $\Lambda$CDM spectrum using the same Planck~2018 binning intervals in $\ell$. The total $\Delta\chi^2$ for both cases is negligible compared to the number of parameters we have.

\section{Shape Analysis for the $\eta_* = 50$~Mpc Signal  }\label{app.50shape}

\begin{figure}[!h]
    \begin{center}
    \includegraphics[width=0.35\textwidth,clip]{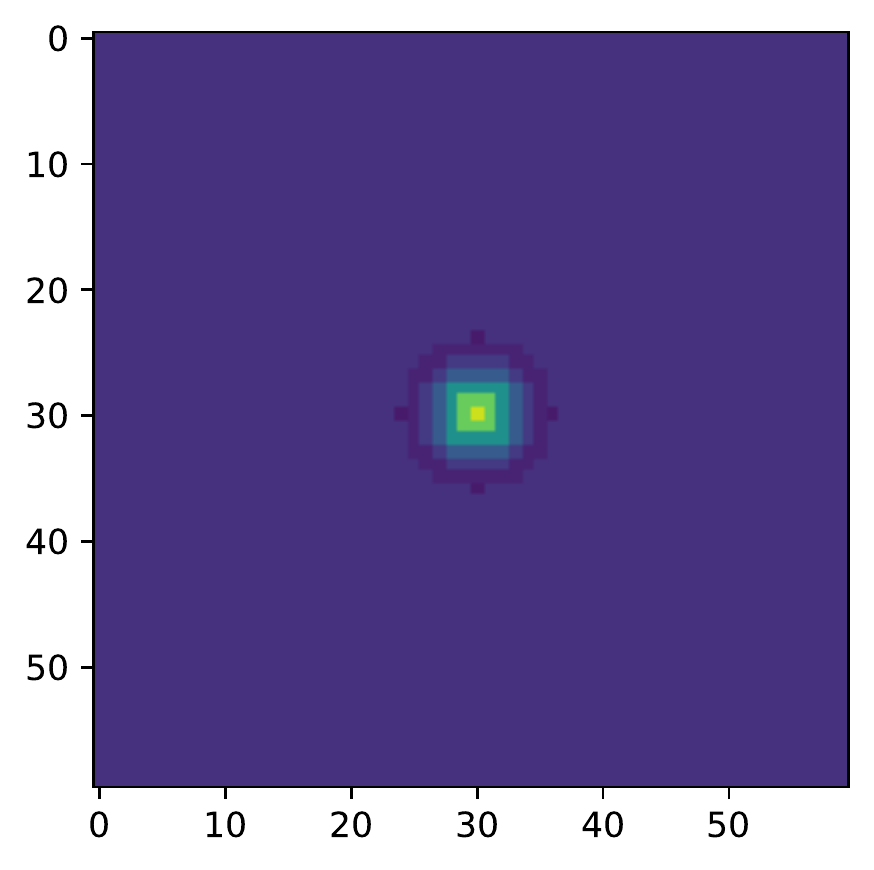}
    \hspace{0.1cm}
    \includegraphics[width=0.6\textwidth,clip]{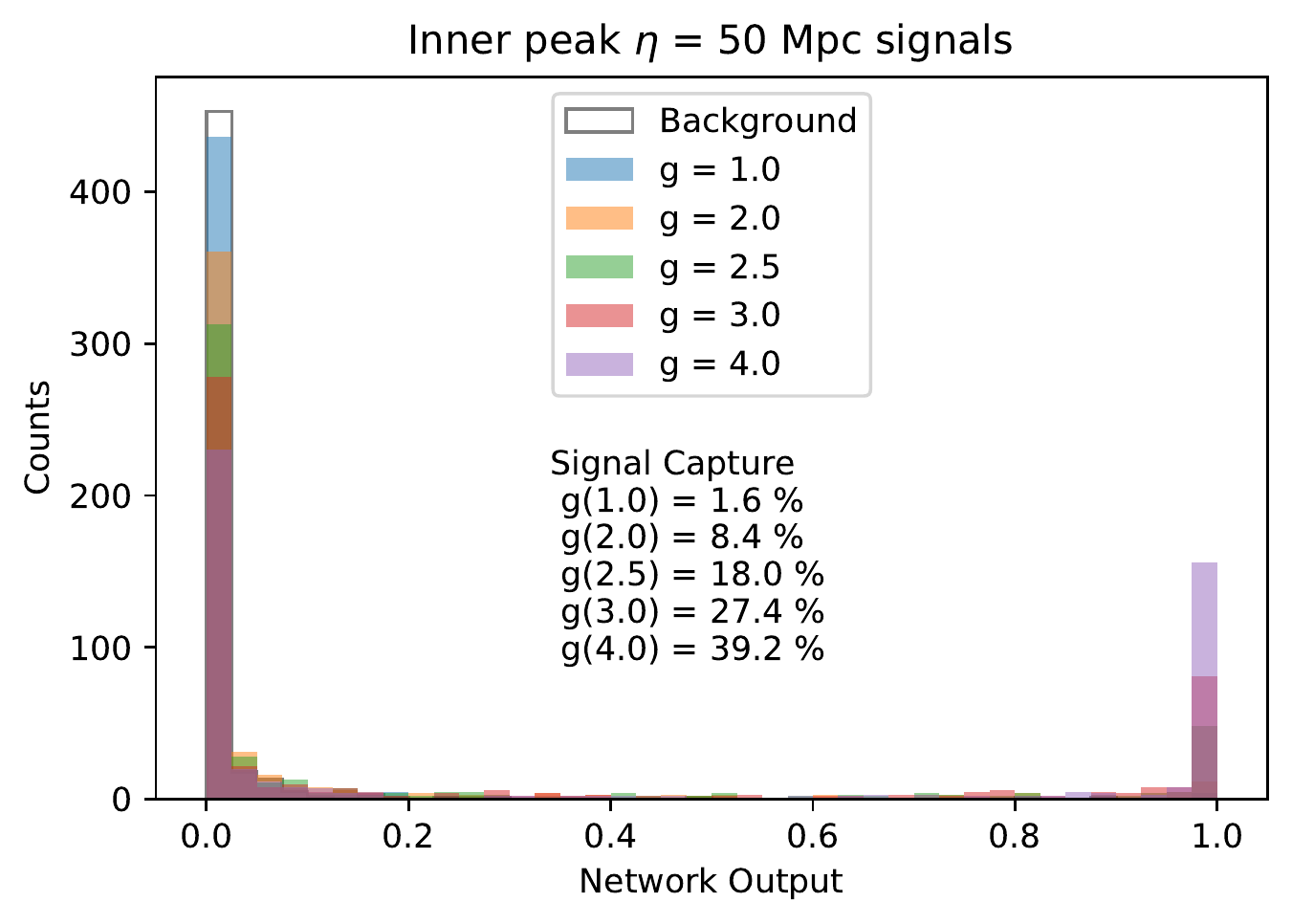}
    \end{center}
    \vspace{-0.1cm}
    \caption{\label{fig:eta50intest} In the left panel we show the trimmed inner piece of a  hotspot signal, while in the right we show the output after 500 CMB + inner hotspot images are run through a network trained on full (untrimmed) $\eta_* = 50\, \text{Mpc}, g = 3$ hotspots.}
\end{figure}

\begin{figure}[!ht]
    \begin{center}
    \includegraphics[width=0.35\textwidth,clip]{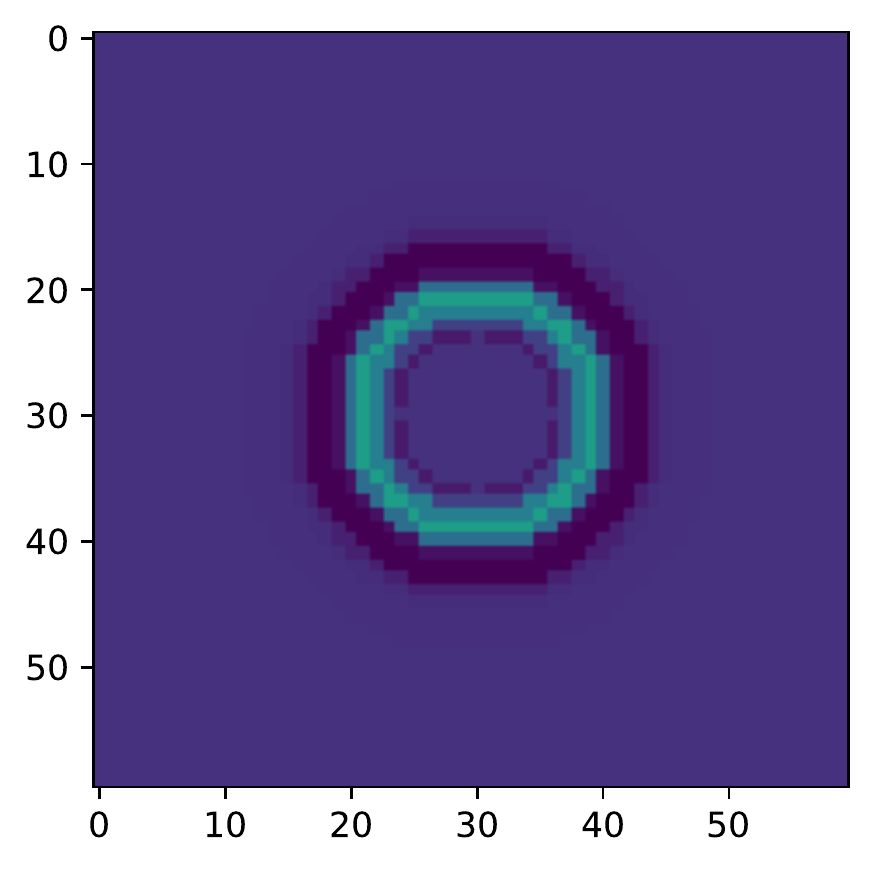}
    \hspace{0.1cm}
    \includegraphics[width=0.6\textwidth,clip]{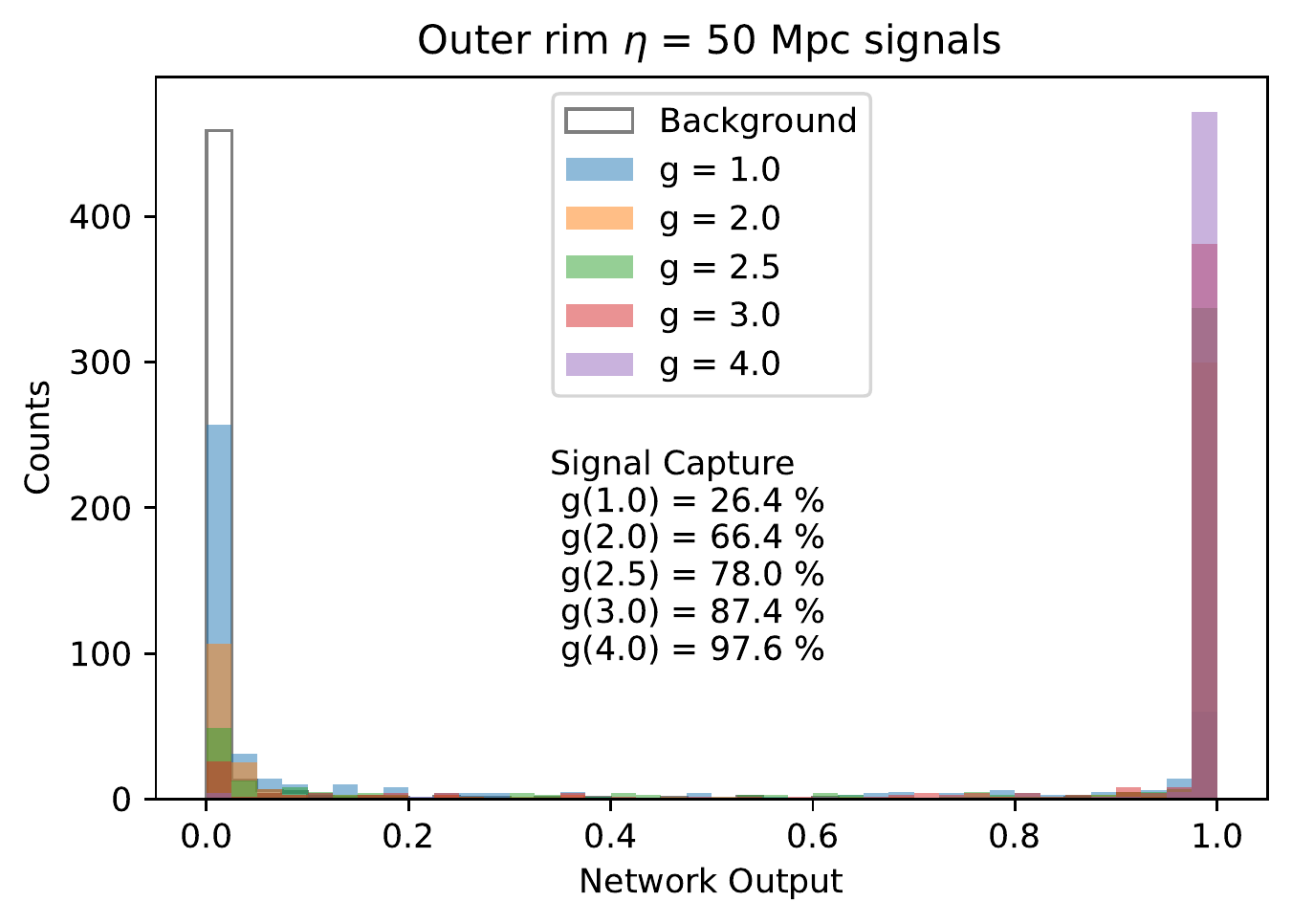}
    \end{center}
    \vspace{-0.1cm}
    \caption{\label{fig:eta50outtest}
   In the left panel we show the trimmed outer piece of a  hotspot signal, while in the right we show the output after 500 CMB + outer hotspot images are run through the same network as in Fig.~\ref{fig:eta50intest}. The capture rates for the ring are much higher than for the central spot, shown in Fig.~\ref{fig:eta50intest}.}
\end{figure}

In our earlier results, we found that the CNN's performance for $\eta=50$~Mpc PHS exceeds the other benchmarks, despite the fact that the hotspots at $\eta=50$~Mpc are much cooler. We surmise that the result is due to the distinct shape of the profile --  a rim structure with central peak. As a simple test of this hypothesis, we formed a signal set of PHS decomposed into two separate features, an inner peak and an outer rim. We then ran each piece through a network trained on the complete shape of the $\eta = 50$ Mpc spots. 

We ran 500 CMB + deconstructed PHS test samples through the network, using a variety of $g$ values but always with both located on the last scattering surface. The results, along with sample images of the deconstructed signals, are shown in Figs. \ref{fig:eta50intest} and \ref{fig:eta50outtest}. Comparing the right hand panels in Figs. \ref{fig:eta50intest} and \ref{fig:eta50outtest}, we see that the network is much more efficient at capturing the ring portion, e.g. $88\%$ capture for $g=3$ compared to $27\%$ for the central spot. From this test we conclude that the ring shape is crucial to the CNN's performance at low $\eta_*$ (note that the signal capture for the ring  nearly matches the capture rate for the full signal (Fig.~\ref{fig:output_distrb})).

\bibliographystyle{JHEP}
\bibliography{bibliography}

\end{document}